\PassOptionsToPackage{svgnames,table}{xcolor}
\documentclass[sigconf]{acmart}

\setcopyright{none}

\usepackage{amsmath}
\usepackage{amssymb}
\usepackage{blindtext}
\usepackage[normalem]{ulem}

\usepackage{listings}
\usepackage{hyperref}
\usepackage{graphicx}
\usepackage{booktabs}  
\usepackage{array}
\usepackage{subcaption}
\usepackage[normalem]{ulem}

\newcommand{\g}[1]{{\textcolor{gray}{#1}}}
\newcommand{\x}[1]{{\uline{#1}}}

\usepackage[T1]{fontenc}  
\usepackage[scaled=0.80]{beramono}  
\usepackage{microtype}
\lstdefinelanguage{scala}{
  alsoletter={@=>},
  morekeywords={
  abstract, case, catch, class, def, do, else, extends, false, final, finally,
  for, if, implicit, import, match, new, null, object, override, package,
  private, protected, requires, return, sealed, super, this, throw, trait, try,
  true, type, val, var, while, with, yield, domain, postcondition,
  precondition,invariant, constraint, assert, forAll,  _, return, @generator,
  ensure, require, ensuring,=>, Real, certainly, possibly, certify, errorBound,
  assertBound
  },
  sensitive=true,
  morecomment=[l]{//},
  morecomment=[s]{/*}{*/},
  commentstyle=\color{gray},
  showstringspaces=false,
  columns=fullflexible,
  mathescape=true,
  numberstyle=\tiny,
  basicstyle=\small\ttfamily,
  numbersep=5pt,
  stepnumber=2,
  numbers=none,                   
  morestring=[b]"
}
\newcommand{\code}[1] {\lstinline!#1!}

\newcommand{\daisy}{Anton} 

\begin{document}
\title{Sound Mixed-Precision Optimization with Rewriting}

\author{Eva Darulova}\affiliation{\institution{MPI-SWS}}\email{eva@mpi-sws.org}
\author{Einar Horn}\affiliation{\institution{UIUC}}\email{eahorn2@illinois.edu}
\author{Saksham Sharma}\affiliation{\institution{IIT Kanpur}}\email{sakshams@cse.iitk.ac.in}

\begin{abstract}

Finite-precision arithmetic computations face an inherent tradeoff between
accuracy and efficiency. The points in this tradeoff space are determined, among
other factors, by different data types but also evaluation orders. To put it
simply, the shorter a precision's bit-length, the larger the roundoff error
will be, but the faster the program will run. Similarly, the fewer arithmetic
operations the program performs, the faster it will run; however, the effect on
the roundoff error is less clear-cut.
Manually optimizing the efficiency of finite-precision programs while ensuring
that results remain accurate enough is challenging. The unintuitive and discrete
nature of finite-precision makes estimation of roundoff errors difficult; furthermore
the space of possible data types and evaluation orders is prohibitively large.

We present the first fully automated and sound technique and tool for optimizing
the performance of floating-point and fixed-point arithmetic kernels. Our
technique \emph{combines} rewriting and mixed-precision tuning. Rewriting
searches through different evaluation orders to find one which minimizes the
roundoff error at no additional runtime cost. Mixed-precision tuning assigns
different finite precisions to different variables and operations and thus
provides finer-grained control than uniform precision. We show that when these
two techniques are designed and applied together, they can provide higher
performance improvements than each alone.

\end{abstract}

%
%



\maketitle

\section{Introduction}

  Finite-precision computations, such as those found in embedded and scientific computing
  applications, face an inherent tradeoff between accuracy and
  efficiency due to unavoidable roundoff errors whose magnitude depends on
  several aspects. One of these is the data type chosen: in general, the larger
  the data type (e.g. in terms of bits), the smaller the roundoff errors will be.
  However, increasing the bit-length typically leads to decreases in
  performance.
  Additionally, finite-precision arithmetic is not associative or distributive.
  Thus, an attempt to reduce the running time of a computation by reducing the
  number of arithmetic operations (e.g. $a * b + a * c \rightarrow a * (b +
  c)$) may lead to a higher and possibly unacceptable roundoff error. Due to
  the unintuitive nature of finite-precision and the subtle
  interactions between accuracy and efficiency, manual optimization
  is challenging and automated tool support is needed.

  \paragraph{Mixed-precision Tuning}

    In order to save valuable resources like time, memory or energy, we would
    like to choose the smallest data type which still provides sufficient accuracy.
    Not all applications require high precision, but how much
    precision an application needs depends highly on context: on the computations
    performed, the magnitude of inputs, and the expectations of the environment,
    so that no one-size-fits-all solution exists. Today, the common way to program
    is to pick a seemingly safe, but often overprovisioned, data type ---
    for instance, uniform double floating-point precision.

    Mixed-precision, where different operations are performed in potentially different
    precisions, increases the number of points on the accuracy-efficiency
    tradeoff space and thus increases the possibility for more resource savings.
    With uniform precision, if one precision is just barely not enough, we are
    forced to upgrade all operations to the next higher precision.
    This can increase the running time of the program substantially. Therefore,
    it would be highly desirable to upgrade only part of the operations; just
    enough to meet the accuracy target, while increasing the execution time by
    the minimum.

    One of the challenges in choosing a finite precision -- uniform or mixed --
    is ensuring that the roundoff errors remain below an
    application-dependent acceptable bound. Recent work has provided automated
    techniques and tools which help the programmer choose between different
    uniform precisions by computing sound worst-case numerical error
    bounds~\cite{Goubault2011,Darulova2014,Solovyev2015,Magron2015}.

    However, selecting a suitable mixed precision is significantly more
    difficult than choosing a uniform precision. The number of
    different type assignments to variables and operations is too large to
    explore exhaustively. Furthermore, neither roundoff errors nor the
    performance of mixed-precision programs follow intuitive rules, making
    manual tuning very challenging. For instance, changing one particular
    operation to lower precision may produce a smaller roundoff error than
    changing two (other) operations. Furthermore, mixed-precision introduces
    cast operations, which may increase the running time, even though the
    accuracy decreases. %

    In high-performance computing (HPC), mixed-precision
    tuning~\cite{Lam2016,Rubio-Gonzalez2013} approaches use dynamic techniques
    to estimate roundoff errors, and thus do not provide accuracy guarantees.
    This makes them unsuitable, for instance, for many safety-critical
    embedded applications.
    The FPTuner tool~\cite{Chiang2017} is able to soundly tune mixed-precision
    for straight-line programs, but it requires user
    guidance for choosing which mixed-precision variants are more efficient and
    is thus not fully automated. Furthermore, its tuning time can be
    prohibitively large.


  \paragraph{Rewriting}

    Another possibility to improve the efficiency of finite-precision arithmetic
    is to reduce the number of operations that need to be carried out. This can
    be achieved without changing the \emph{real-valued} semantics of the program
    by rewriting the computation using laws like distributivity and
    associativity. Unfortunately, these laws do not hold for finite-precision
    computations: changing the order of a computation changes the magnitude of
    roundoff errors committed, but in unintuitive ways.
    Previous work has focused on automated techniques for finding a rewriting
    (i.e. re-ordering of computations) such that roundoff errors are
    minimized~\cite{Darulova2013,Panchekha2015}. However, optimizing for
    accuracy may increase the number of arithmetic operations and thus the
    execution time.



  \paragraph{Combining Mixed-Precision Tuning and Rewriting}

    We propose to combine mixed-precision tuning with rewriting in a
    \emph{fully automated} technique for performance optimization of
    finite-precision arithmetic computations. Our approach is \emph{sound} in
    that generated programs are guaranteed to satisfy user-specified roundoff
    error bounds, while our performance improvements are best effort (due
    to the complexity and limited predictability of today's hardware).
    Our rewriting procedure takes into account both accuracy and the number of
    arithmetic operations. It can reduce the running time of programs
    directly, but more importantly, by
    improving the accuracy, it allows for more aggressive mixed-precision tuning.
    To the best of our knowledge, this is the first combination of
    mixed-precision tuning and rewriting for performance optimization.

    We combine a search algorithm which was successfully applied in HPC
    mixed-precision tuning~\cite{Rubio-Gonzalez2013} with a sound static
    roundoff error analysis~\cite{Darulova2014} and a static performance cost
    function to obtain a mixed-precision tuning technique which is both sound
    and fully automated as well as efficient. We furthermore modify a rewriting
    optimization algorithm based on genetic programming~\cite{Darulova2013} to
    consider both accuracy and performance.

    While most of the building blocks of our approach have been presented
    before, their effective combination requires careful adaptation and
    coordination, as our many less-successful experiments have shown. Just as a
    manual optimization is challenging due to the subtle interactions of
    finite-precision accuracy and performance, so is the design of an automated
    technique.

    We focus on arithmetic kernels, and do not consider conditionals and loops.
    Our technique (as well as FPTuner's) can be extended to conditionals by
    considering individual paths separately as well as to loops by optimizing
    the loop body only and thus reducing both to straight-line code. The challenge
    currently lies in the sound roundoff error estimation, which is known to be
    hard and expensive for conditionals and
    loops~\cite{Darulova2017,Goubault2013}, and is largely orthogonal to the
    focus of this paper.

    Our technique is applicable and implemented in a tool called \daisy{} for
    both floating-point as well as fixed-point arithmetic. While floating-point
    support is standardized, fixed-point arithmetic is most effective in
    combination with specialized hardware. We focus in this paper on the
    algorithmic aspects of optimizing arithmetic kernels and leave a thorough
    investigation of specialized hardware implementations for future work.

    For floating-point arithmetic, we evaluate \daisy{} on standard benchmarks from embedded
    systems and scientific computing.
    We observe that rewriting alone improves performance by up to 17\% and
    for some benchmarks even more by reducing roundoff errors sufficiently
    to enable uniform double precision where the original benchmark requires
    (uniform) quad precision.
    Mixed-precision tuning improves performance by up to 45\% when compared
    to the uniform precision version which would be needed to satisfy the required
    accuracy.
    In combination with rewriting, \daisy{} improves performance by up to 54\%
    (93\% for those cases where rewriting enables uniform double precision),
    and we also observe that it improves performance for more benchmarks
    than when using mixed-precision tuning or rewriting alone.

    We focus on performance, although our algorithm is independent of the optimization
    objective such that - with the corresponding cost function - memory or energy
    optimization is equally possible.

  \paragraph{Contributions}
    To summarize, in this paper we present:

    \begin{itemize}
      \item An optimization procedure based on rewriting, which takes into
      account both accuracy and performance.

      \item A sound, fully automated and efficient mixed-precision
      tuning technique.

      \item A carefully designed combination of rewriting and mixed-precision
      tuning, which provides more significant performance
      improvements that each of them alone.


      \item An implementation in a tool called \daisy{}, which
      generates optimized source programs in Scala and in C and which supports both
      floating-point as well as fixed-point arithmetic. We plan to release
      \daisy{} as open source.

      \item We show the effectivness of our tool on a set of arithmetic kernels
      from embedded systems and scientific computing and compare it against
      the state-of-the-art.

    \end{itemize}

\section{Overview}\label{sec:overview}
  We first provide a high-level overview and explain the key ideas of our approach
  using an example.
  Inspired by the tool Rosa~\cite{Darulova2014}, the input to our tool \daisy{}
  is a program written in a real-valued specification language. (Nothing in our technique
  depends on this particular frontend though.)
  Each program consists of a number of functions which are optimized
  separately. The following nonlinear embedded controller~\cite{Darulova2013} is one such example function:
  %
  \begin{lstlisting}
  def rigidBody1(x1: Real, x2: Real, x3: Real): Real = {
    require(-15.0 <= x1 && x1 <= 15 && -15.0 <= x2 &&
        x2 <= 15.0 && -15.0 <= x3 && x3 <= 15)
    -x1*x2 - 2*x2*x3 - x1 - x3
  } ensuring(res => res +/- 1.75e-13)
  \end{lstlisting}
  %
  In the function's precondition (the \lstinline{require} clause) the user
  provides the ranges of all input variables, on which the magnitudes of roundoff errors depend.
  The postcondition (the \lstinline{ensuring} clause) specifies the required accuracy of the result
  in terms of worst-case absolute roundoff error. For our controller, this
  information may be, e.g., determined from the specification of the system's sensors
  as well as the analysis of the controller's stability~\cite{Majumdar2012}.
  The function body consists of an arithmetic expression (with $+, -, *, /, \sqrt{}$) with possibly local
  variable declarations.

  As output, \daisy{} generates a mixed-precision source-language program,
  including all type casts, which is guaranteed to satisfy the given error bound and is
  expected to be the most efficient one among the possible candidates.
  \daisy{} currently supports fixed-point arithmetic with bitlengths of 16 or 32
  bits or IEEE754 single (32 bit) and double (64 bit) floating-point precision
  as well as quad precision (128 bit). The latter can be implemented on top of regular
  double-precision floating-points~\cite{Bailey2015}. Anton can be easily
  extended to support other fixed- or floating-point precisions; here we have
  merely chosen a representative subset.




  Our approach decouples rewriting from
  the mixed-precision tuning. To find the optimal program, i.e. the most
  efficient one given the error constraint, we would need to optimize both the
  evaluation order as well as mixed-precision simultaneously:
  i) the evaluation order determines which mixed-precision type assignments to
  variables and operations are feasible, and
  ii) the mixed-precision assignment influences which evaluation order is optimal.
  Unfortunately, this would require an exhaustive search~\cite{Darulova2013},
  which is computationally infeasible. We thus choose to separate rewriting from
  mixed-precision tuning and further choose (different) efficient search
  techniques for each.


  \paragraph{Step 1: Rewriting}

    \daisy{} first rewrites the input expression into one which is
    equivalent under a real-valued semantics, but one which has a smaller
    roundoff error when implemented in finite-precision and which does not
    increase the number of arithmetic operations. Rewriting can increase the
    opportunities for mixed-precision tuning, because a smaller roundoff error
    may allow more lower-precision operations. The second objective makes sure
    that we do not accidentally increase the execution time of the program by
    performing more arithmetic operations and even lets us improve the
    performance of the expression directly.

    \daisy{}'s rewriting uses a genetic algorithm to search the vast space
    of possible evaluation orders efficiently. At every iteration, the algorithm
    applies real-valued identities, such as associativity and distributivity,
    to explore different evaluation orders. The search is guided by a fitness
    function which bounds the roundoff errors for a candidate expression -
    the smaller the error, the better. This error computation is done wrt.
    uniform precision, as the mixed-precision type assignment will only be determined
    later.
    While the precision can affect the result of rewriting, we empirically show
    that the effect is small (\autoref{sec:rewriting}).


    This approach is heavily inspired by the algorithm presented
    in~\citeauthor{Darulova2013}~\cite{Darulova2013} which optimized for
    accuracy only. We have made important adaptations, however, to make it work in
    practice for optimizing for performance as well as to work well with
    mixed-precision tuning.

    For our running example, the rewriting phase produces the following
    expression, which improves accuracy by 30.39\% and does not change the
    number of operations:
    \begin{lstlisting}
    (-(x1 * x2) - (x1 + x3)) - ((2.0 * x2) * x3)
    \end{lstlisting}
    To give an intuition why this seemingly small change makes such a
    difference, note that the magnitude of roundoff errors depends on the possible ranges of
    intermediate variables. Even small changes in the evaluation order can have
    large effects on these ranges and consequently also on the roundoff errors.

  \paragraph{Step 2: Code Transformation}
    To facilitate mixed-precision tuning, \daisy{} performs two code
    transformations: constants are assigned to fresh variables and the remaining
    code is converted into three-address form. By this, every constant and
    arithmetic operation corresponds to exactly one variable, whose precision
    will be tuned during phase 4. If not all arithmetic operations should be
    tuned, i.e. a more coarse grained mixed-precision is desired, then this
    step can be skipped.

  \paragraph{Step 3: Range Analysis}
    The evaluation order now being fixed, \daisy{} computes the real-valued ranges of all
    intermediate subexpressions and caches the results. Ranges are needed
    for bounding roundoff errors during the subsequent mixed-precision tuning,
    but because the \emph{real-valued} ranges are not affected by different
    precisions, \daisy{} computes them only once for efficiency.
    %

  \paragraph{Step 4: Mixed-precision Tuning}
    To effectively search the space of possible mixed-precision type
    assignments, we choose a variation of the delta-debugging algorithm
    used by Precimonious~\cite{Rubio-Gonzalez2013}, which prunes the
    search space in an effective way. It starts with all variables in the highest available precision
    and attempts to lower variables in a systematic way until it finds that no
    further lowering is possible while still satisfying the given error bound.
    We have also tried to apply a genetic algorithm for mixed-precision tuning,
    but observed that it was quite clearly not a good fit (we omit the
    experimental results for space reasons).

    Unlike Precimonious, which evaluates the accuracy and performance of
    different mixed-precisions dynamically, \daisy{} uses a static sound error
    analysis as well as a static (but heuristic) performance cost function to
    guide the search. The performance cost function assigns (potentially
    different) abstract costs to each arithmetic as well as cast operation.
    Using static error and cost functions reduces the tuning time significantly,
    and further allows tuning to be run on different hardware than the final
    generated code.

    For our running example, \daisy{} determines that uniform double floating-point precision
    is not sufficient and generates a tuned program which runs 43\% faster than the quad
    uniform precision version, which is the next available uniform precision in the
    absence of mixed-precision:
    \begin{lstlisting}[mathescape=true]
  def rigidBody1(x1: Quad, x2: Quad, x3: Double): Double = {
    ($-_{d}$(x1 $*_{q}$ x2) $-_{d}$ (x1 $+_{q}$ x3)) $-_{d}$ ((x2 $*_{q}$ 2.0f) $*_{d}$ x3)
  }
  \end{lstlisting}
    For readability, we have inlined the expression and use the letters `d' and
    `q' to mean that the operation is performed in double and quad precision
    respectively.
    The entire optimization including rewriting takes about 4 seconds.
    Had we used only the mixed-precision tuning without rewriting, the program
    would still run 28\% faster than quad precision.

  \paragraph{Step 5: Code Generation}
    Once mixed-precision tuning finds a suitable type configuration, \daisy{}
    generates the corresponding finite-precision program (in Scala or C), inserting all
    necessary casts, and in the case of fixed-point arithmetic all necessary
    bit-shift operations.

    Our entire approach is parametric in the finite-precision used, and thus
    works equally for fixed-point arithmetic.
    %
    Furthermore, it is geared towards optimizing the performance of programs under
    the (hard) constraint that the given error bound is guaranteed to be
    satisfied. Other optimization criteria like memory and energy are also
    conceivable, and would only require changing the cost function.

\section{Background}\label{sec:background}

We first review necessary background about finite-precision arithmetic and
sound roundoff error estimation, which is an important building block
for both rewriting and mixed-precision tuning.

\subsection{Floating-point Arithmetic}
  We assume standard IEEE754 single and double precision floating-point
  arithmetic, in rounding-to-nearest mode and the following standard abstraction
  of IEEE754 arithmetic operations:\\
  $
  \label{eqn:floats}
    x \circ_{fl} y = (x \circ y)(1 + \delta) \;\text{, }\; \lvert \delta \rvert \le \epsilon_m
  $
  where $\circ \in {+, -, *, /}$ and $\circ_{fl}$ denotes the respective
  floating-point version, and $\epsilon_m$ bounds the maximum relative error
  (which is $2^{-24}$, $2^{-53}$ and $2^{-113}$ for single, double and quad
  precision respectively). Unary minus and square root follow similarly.
  We further consider NaNs (not-a-number special values), infinities and ranges
  containing only denormal floating-point numbers to be errors and \daisy{}'s
  error computation technique detects these automatically. We note that under
  these assumptions the abstraction is indeed sound.

\subsection{Fixed-point Arithmetic}

  Floating-point arithmetic requires dedicated support, either in hardware or in
  software, and depending on the application, this support may be too costly. An
  alternative is fixed-point arithmetic which can be implemented with integers
  only, but which in return requires that the radix point alignments are
  precomputed at compile time. While no standard exists, fixed-point values are
  usually represented as (signed) bit vectors with an integer and a fractional
  part, separated by an implicit radix point. At runtime, the alignments are
  then performed by bit-shift operations. These shift operations can also be
  handled by special language extensions for fixed-point
  arithmetic~\cite{ISOIEC2008}. For more details please see~\citeN{Anta2010},
  whose fixed-point semantics we follow. We use truncation as the rounding mode
  for arithmetic operations. The absolute roundoff error at each operation is
  determined by the fixed-point format, i.e. the (implicit) number of fractional
  bits available, which in turn can be computed from the range of possible
  values at that operation.

\subsection{Sound Roundoff Error Estimation}\label{sec:error-analysis}
  We build upon Rosa's static error computation~\cite{Darulova2014},
  which we review here.
  Keeping with Rosa's notation, we denote by $f$ and $x$ a mathematical
  real-valued arithmetic expression and variable, respectively, and by
  $\tilde{f}$ and $\tilde{x}$ their floating-point counterparts. The
  worst-case absolute error that the error computation approximates is
  $
  \label{eqn:absError}
  \max_{x \in [a, b]}\;\;\lvert f(x) - \tilde{f}(\tilde{x}) \rvert
  $
  where $[a, b]$ is the range for $x$ given in the precondition. The input $x$ may
  not be representable in finite-precision arithmetic, and thus we consider an
  initial roundoff error:
  $\lvert x - \tilde{x} \rvert = \lvert x \rvert * \delta,\;\; \delta \le \epsilon_m$
  which follows from~\autoref{eqn:floats}. This definition extends to
  multi-variate $f$ component-wise.

  At a high level, error bounds are computed by a data-flow analysis over the
  abstract syntax tree, which computes for each intermediate arithmetic
  expression (1) a bound on the real-valued range, (2) using this range, the
  propagated errors from subexpressions and the newly committed worst-case
  roundoff error. For a more detailed description, please
  see~\cite{Darulova2014}.

  For our rewriting procedure, since intermediate ranges change
  with different evaluation orders, we compute both the ranges and the errors at
  the same time. For mixed-precision tuning, where real-valued ranges remain
  constant, we separate the computations and only compute ranges once.

  \daisy{} currently does not support additional input errors, e.g. from noisy sensors,
  but note that an extension is straight-forward. In fact,
  a separate treatment of roundoff errors and propagated input errors may be
  desirable~\cite{Darulova2017}.

  We compute absolute errors. An automated and general estimation of relative
  errors ($\lvert f(x) - \tilde{f}(\tilde{x})
  \rvert / \lvert f(x)\rvert$), though it may be more desirable, presents a
  significant challenge today. To the best of our knowledge, state-of-the-art
  static analyses only compute relative errors from an absolute error bound, which
  is then not more informative.
  Furthermore, relative error is only defined if the range of the expression in
  question (i.e. the range of $f(x)$) does not include zero, which unfortunately
  happens very often in practice.

  %
  \paragraph{Range Estimation}
  Clearly, accurately estimating ranges is the main component in the error bound
  computation, and is known to be challenging, especially for nonlinear
  arithmetic. This challenge was addressed in
  previous work on finite-precision verification~\cite{Darulova2014,Goubault2011,Solovyev2015}.
  Interval arithmetic (IA)~\cite{Moore1966} is an efficient choice for range
  estimation, but one which often introduces large over-approximations as it
  does not consider correlations between variables. Affine
  arithmetic~\cite{Figueiredo2004} tracks \emph{linear} correlations, and is
  thus sometimes better (though not always) in practice.
  The overapproximations due to nonlinear arithmetic ($*, /, \sqrt{}$) can be
  mitigated by refining ranges computed by IA with a complete (though expensive)
  nonlinear arithmetic decision procedure inside the Z3~\cite{De-Moura2008} SMT
  solver~\cite{Darulova2014}. Anton's computation builds on this work and is
  parametric in the range arithmetic and currently supports interval and affine
  arithmetic as well as the combination with SMT.



\section{Rewriting Optimization}\label{sec:rewriting}

  The goal of \daisy{}'s rewriting optimization is to
  find an order of computation which is equivalent to the original expression
  under a real-valued semantics, but which exhibits a smaller roundoff error in
  finite-precision - while not increasing the execution time. We first review
  previous work that we build upon and then describe the concrete adaptation in
  \daisy{}.

  \subsection{Genetic Search for Rewriting}\label{sec:genetic-rewriting}
    An exhaustive search of all possible rewritings or evaluation orders is
    computationally infeasible. Even for only linear arithmetic, the problem of
    finding an optimal order is NP-hard~\cite{Darulova2013} and does not allow a
    divide-and-conquer or gradient-based method such that a heuristic and
    incomplete search becomes necessary.

    Genetic programming~\cite{Poli08} is an evolutionary heuristic search
    algorithm which iteratively evolves (i.e. improves) a population of
    candidate expressions, guided by a fitness function. The search is
    initialized with copies of the initial expression. At every iteration,
    expressions are selected from the current population based on their fitness,
    and then mutated to form the the next population. For rewriting, the fitness
    is the worst-case roundoff error - the smaller the better. The selected
    expressions are randomly mutated using mathematical real-valued identities,
    e.g. $a + (b + c) \to (a + b) + c$.
    In this fashion, the algorithm explores different rewritings. The key idea
    is that the likelyhood of an expression to be selected depends on its
    fitness - fitter expressions are more likely to be selected - and thus the
    search converges with each iteration towards expressions with smaller
    roundoff errors. Furthermore, even less-fit expressions have a non-zero
    probability of being selected, thus helping to avoid local minima.

    The output of the procedure is the expression with the least roundoff error
    seen during the run of the search.
    \citeauthor{Darulova2013}~\cite{Darulova2013} used a static analysis as
    described in~\autoref{sec:error-analysis} as the fitness function, with
    smaller roundoff error being better. Their approach was implemented for
    fixed-point arithmetic only, and the optimization goal was to reduce
    roundoff errors.

  \subsection{Rewriting in \daisy{}}

    We instantiate the algorithm described above with a population of 30
    expressions, 30 iterations and tournament selection~\cite{Poli08} for
    selecting which expressions to mutate. These are successfull settings
    identified in~\cite{Darulova2013}. We do not use the
    crossover operation, because it had only limited effects.
    We further extend the rather limited set of mutation rules
    in~\cite{Darulova2013} with the more complete one used by the (unsound) rewriting optimization
    tool called Herbie~\cite{Panchekha2015} (see~\autoref{sec:related}).
    These rules are still based on mathematical identities.
    For the static error function, as described in~\autoref{sec:error-analysis}, we choose
    interval arithmetic for computing ranges and affine arithmetic for tracking
    errors, which provide a good accuracy-performance tradeoff.

    The algorithm described in~\autoref{sec:genetic-rewriting} reduces
    roundoff errors, but may - and as we experimentally observed often does -
    increase the number of operations and thus the execution time. This may
    negate any advantage reduced roundoff errors bring for mixed-precision
    tuning. Furthermore, it is not clear at this point with respect to which
    precision to perform the rewriting as a mixed-precision type assignment is
    not available.

    \subsubsection{Optimizing for Performance}

     We modify the search algorithm to return the expression which does not
     increase the number of arithmetic operations beyond the initial count, and
     which has the smallest worst-case roundoff error.
     We do not use a more sophisticated cost function, as for this the actual
     final type assignment would be needed (which only becomes available after
     mixed-precision tuning).
     We have also implemented a variation of the search which minimizes the
     number of arithmetic expressions, while not increasing the roundoff error
     beyond the error of the initial expression. However, we found empirically
     in our experiments that this produces worse overall results in combination
     with mixed-precision tuning, i.e. reducing the roundoff was more beneficial
     than reducing the operation count. For space reasons, we omit this
     experimental comparison.

    \subsubsection{Optimizing with Uniform Precision}

      The static error analysis, which we use as the fitness function during
      search has to be performed wrt. to some mixed or uniform precision,
      and different choices may result in the algorithm returning syntactically
      different rewritten expressions.
      As the final (ideal) mixed-precision type assignment is not available
      when \daisy{} performs rewriting, it has to choose some precision without
      knowing the final assignment.

      The main aspect which determines which evaluation order is better over
      another are the ranges of intermediate variables - the larger the ranges,
      the more already accumulated roundoff errors will be magnified. These
      intermediate ranges differ only little between different precisions,
      because the roundoff errors are small in comparison to the real-valued
      ranges. 
      Thus, we do not expect that different precision affect the result of
      rewriting very much.

      We performed the following experiment to validate this intuition. We ran
      rewriting repeatedly on the same expression, but with the error analysis
      wrt. uniform single, double and quad floating-point precision as well as up
      to 50 random mixed-precision type assignments. We picked each
      mixed-precision assignment as the baseline in turn. We evaluated the
      roundoff errors of the expressions returned by the uniform precision
      rewritings under this mixed-precision assignment. If rewriting in uniform
      precision produces an expression which has the same or a similar error as
      the expressions returned with rewriting wrt. mixed-precision, then we consider the
      uniform precision to be a good proxy. We counted how often each of the three
      uniform precisions was such a good proxy, where we chose the threshold to
      be that the errors should be within 10\%.
      For space reasons, we only summarize the results. Single and double
      floating-point precision were a good proxy in roughly 80\% of the cases,
      whereas quad precision in 75\%. When the mixed-precision assignments
      were in fixed-point precision, single, double and quad uniform
      precision all achieve 69\% accuracy. Performing the rewriting wrt.
      fixed-point arithmetic is not more beneficial either. Finally, rewriting in uniform
      precision never increased the errors (when evaluated in the mixed-precision baseline).
      We thus (randomly) choose to perform the rewriting with respect to double
      floating-point precision.


\section{Sound Mixed-Precision Tuning}\label{sec:mixed}

  After rewriting, \daisy{} pre-processes expressions as described in step 2
  in~\autoref{sec:overview} and computes the now-constant ranges of intermediate
  expressions. Since the range computation needs to be performed only once, we
  choose the more expensive but also more accurate range analysis using a
  combination of interval arithmetic and SMT~\cite{Darulova2014}. We again first
  review previous work that we build upon before explaining \daisy{}'s technique
  in detail.

\subsection{Delta-Debugging for Precision Tuning}\label{sec:delta-debugging}

\begin{figure}
    \centering
    \includegraphics[width=0.30\textwidth]{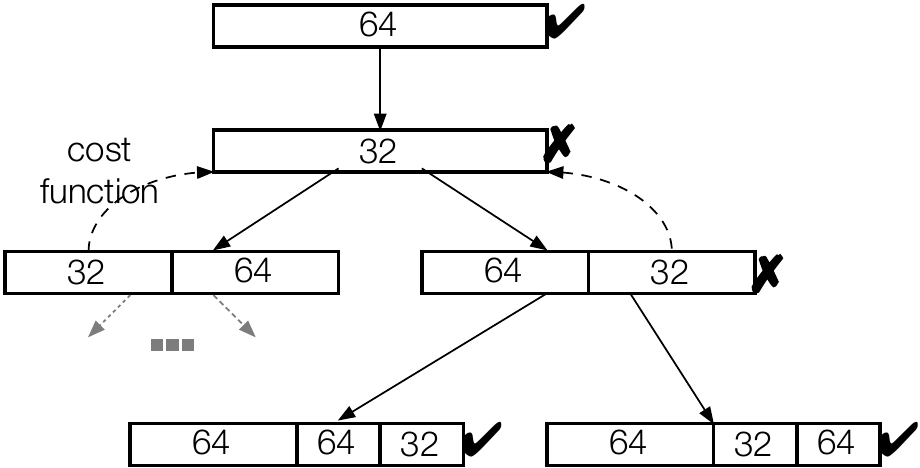}
    \caption{Sketch of the delta-debugging algorithm}
    \label{fig:delta-debugging}
    \vspace{-0.5cm}
  \end{figure}

  Delta-debugging has been originally conceived in the context of software testing for
  identifying the smallest failing testcase~\cite{Zeller2002}.
  \citeauthor{Rubio-Gonzalez2013}~\cite{Rubio-Gonzalez2013} have adapted this
  algorithm for mixed-precision tuning in the tool Precimonious.
  %
  It takes as input:
  \begin{itemize}
    \item a list of variables to be tuned ($\tau$) as well as a list of all other variables
    with their constant precision assignments ($\phi$)
    \item an \emph{error function} which bounds the roundoff error
      of a given precision assignment
    \item a \emph{cost function} approximating the expected performance
    \item an error bound $e_{max}$ to be satisfied.
  \end{itemize}
  The output is a precision assignment for variables in $\tau$.
  A partial sketch of the algorithm is depicted in~\autoref{fig:delta-debugging},
  where the boxes represent sets of variables in $\tau$.
  Consider the case where variables can be in single (32 bit) and double (64 bit) floating-point
  precision. The algorithm starts by assigning all variables in $\tau$ to the
  highest precision, i.e. to double precision. It uses the error function to
  check whether the roundoff error is below $e_{max}$. If it is not, then an
  optimization is futile, because even the largest precision cannot satisfy the
  error bound.

  If the check succeeds, the algorithm tries to \emph{lower} all variables in
  $\tau$ by assigning them to single precision. Again, it computes the maximum
  roundoff error. If it is below $e_{max}$, the search stops as single precision is
  sufficient. If the error check does not succeed, the algorithm splits $\tau$
  into two equally sized lists $\tau_1$ and $\tau_2$ and recurses on each separately. When
  recursing on $\tau_1$, the new list of variables considered for lowering
  becomes $\tau' = \tau_1$ and the list of constant variables becomes $\phi' = \phi
  + \tau_2$. The case for recursing on $\tau_2$ is symmetric.
  When a type assignment is found which satisfies the error bound $e_{max}$, the
  recursion stops. Since several valid type assignments can be found, a cost
  function is used to select the one with lowest cost (i.e. best performance.)



  The algorithm is generalized to several precisions by first running it
  with the highest two precisions. In the second iteration, the variables which
  have remained in the highest precision become constant and move to $\phi$. The
  optimization is then performed on the new $\tau$ considering the second and
  third highest precision.

\subsection{Mixed-Precison Tuning in \daisy{}}\label{sec:delta-debug}

  We have instantiated this algorithm in \daisy{} and
  describe now our adaptations which were important to obtain
  \emph{sound} mixed-precision assignments as well as good results.

  \subsubsection{Static Error Analysis}

    Precimonious estimates roundoff errors by dynamically evaluating the program on several
    random inputs. This approach is not sound, and also in
    general inefficient, as a large number of program executions is needed for a
    reasonably confident error bound.
    \daisy{} uses a \emph{sound} \emph{static} roundoff error analysis, which is
    an extension of Rosa's uniform-precision error analysis
    from~\autoref{sec:error-analysis} to support mixed-precision. This extension
    uses affine arithmetic for the error computation and considers roundoff
    errors incurred due to down-casting operations.

    Precimonious can handle any program, including programs with loops.
    Our static error function limits which kinds of programs \daisy{} can handle
    to those for which the error bounds can be verified, but for those it provides
    accuracy \emph{guarantees}. We note,
    however, that our approach can potentially be extended to loops with
    techniques from~\cite{Darulova2017}, by considering the loop body only.


  \subsubsection{Tuning All Variables}

    Unlike Precimonious, which optimizes only the precisions of declared
    variables, \daisy{} optimizes the precisions of all variables and
    intermediate expressions by transforming the program prior to
    mixed-precision tuning into three-address form (this
    transformation can also be skipped, if desired).

    The precision of an arithmetic operation is determined by the precisions of
    its operands as well as the variable that the result is assigned to. In
    general, we follow a standard semantics, where the operation is
    performed in the highest operand precision, with one exception.
    For example, for the precision assignment $\{x \to \text{single}, y \to
    \text{single}, z \to \text{double}\}$, and expression \lstinline{val z = x + y}, we choose
    the interpretation
    \lstinline{z = x.toDouble + y.toDouble} instead of
    \lstinline{z = (x + y).toDouble}, so that the operation is performed in the
    higher precision, thus loosing less accuracy. Our experiments (whose results
    are not shown for space reasons) have confirmed
    that this indeed provides better overall results.


    Delta-debugging operates on a list of variables that it optimizes. We
    have observed in our experiments that it is helpful when the variables are
    sorted by order of appearance in the program. Our hypothesis is that
    delta-debugging is more likely to assign `neighboring' variables the same
    type, which in general is likely to reduce type casts and thus cost.

    We found that often constants are representable in
    the lowest precision, e.g. when they are integers. It is thus tempting to
    keep those constants in the lowest precision. However, we found that,
    probably due to cast operations, this optimization was not a universal
    improvement, so that Anton optimizes constants just like other variables.

  \subsubsection{Static Cost Function}
    Precimonious uses dynamic evaluation to estimate the expected
    running time. We note that this approach is quite inefficient, but also not
    entirely reliable, as running times can vary substantially between runs
    (our benchmarking tool takes several seconds per benchmark until steady-state). It
    furthermore restricts the tuning to the specific platform that the tuning is run on.
    FPTaylor, on the other hand, optimizes for the number of lower-precision
    operations (more is better) and provides a way for the user
    to manually restrict the overall number of cast operations, and
    provides the possibility to constrain certain variables to have the same
    precision (`ganging'). Knowing up front how many cast operations are needed
    is quite challenging.

    We instead propose a static cost function to obtain an overall
    technique which is efficient as well as fully automated. Note that this
    function needs to be able to distinguish only which of two mixed-precision
    assignments is the more efficient one, and does not need to predict the
    actual running times. We are aiming for a practical solution and are aware
    that more specialized approaches are likely to provide
    better prediction accuracy. As we focus in this paper on the algorithmic
    aspects, we leave this for future work.
    We have implemented and experimentally evaluated several cost function candidates
    for floating-point arithmetic, which all require only a few milliseconds to run:
    \begin{itemize}
      \item \emph{Simple cost}
        assigns a cost of 1, 2 and 4 to single, double and quad
        precision arithmetic and cast operations respectively.

      \item \emph{Benchmarked cost}
        assigns abstract costs to each operation based on benchmarked average running
        times, i.e. we benchmark each operation in isolation with random inputs.
        This cost function is platform specific and different arithmetic operations
        have different costs (e.g. addition is generally cheaper than division).

      \item \emph{Operation count cost}
        counts the number of operations performed in each
        precision, recorded as a tuple and ordered lexicographically, i.e.
        more higher-precision operations lead to a higher cost. This cost
        function is inspired by FPTuner and does not
        consider cast operations.

      \item \emph{Absolute errors cost}
        uses the static roundoff error, with smaller values
        representing a higher cost. A smaller roundoff usually
        implies larger data types, which should correlate with a higher
        execution time.
    \end{itemize}

    We evaluate our cost functions experimentally on complete examples (see~\autoref{sec:experiments}).
    For each example function, we first
    generate 42 random precision assignments and their corresponding programs in Scala.
    We calculate the cost of each with all cost functions and also benchmark the
    actual running time. We use Scalameter~\cite{Prokopec2012} for benchmarking
    (see~\autoref{sec:experimental-setup}).
    We are interested in distinguishing pairs of mixed-precision
    assignments, thus for each benchmark program, we create pairs of all the randomly
    generated precision assignments. Then we count how often each static
    cost function can correctly distinguish which of the two assignments is faster, where
    the `ground truth' is given by the benchmarked running times.
    %
    The following table summarizes the results of our cost function evaluation.
    The rows `32 - 128' and `32 - 64' and give the proportion of correctly distinguished
    pairs of type assignments with the random types selected from single, double and quad
    and single and double precision, respectively.
    \begin{center}
    \small
      \renewcommand{\arraystretch}{1.0}
      \begin{tabular}{@{}l|cccc@{}}
        \toprule
        precisions & bench & simple & opCount & errors  \\
        \midrule
        32 - 128 & 0.7692  & \bf{0.8204}  & 0.8106  & 0.5871  \\
        32 - 64  & \bf{0.6416}  & 0.5889  & 0.5477  & 0.5462  \\
        \bottomrule
      \end{tabular}
    \end{center}

    Given these results, we choose a two-pronged approach for floating-point
    arithmetic: whenever quad precision may appear (e.g. during the first round of
    delta-debugging), we use the naive cost function.
    Once no more quad precision appears, we use the benchmarked one (e.g.
    during the second round of delta-debugging for
    benchmarks which do not require quad precision).
    For optimizing fixed-point arithmetic we use the simple cost function.

\section{Implementation and Evaluation}\label{sec:experiments}


    We have implemented \daisy{} in the Scala programming language. Internal arithmetic
    operations are implemented with a rational data type to avoid roundoff errors
    and ensure soundness. Apart from the (optional) use of the Z3 SMT
    solver~\cite{De-Moura2008} for more accurate range computations, \daisy{} does
    not have any external dependencies.


    In this work, we focus on arithmetic kernels, and do not consider
    conditionals and loops. Our technique (as well as FPTuner's) can be extended
    to conditionals by considering individual paths separately as well as to loops by
    optimizing the loop \emph{body} and thus reducing it to straight-line code. The
    challenge currently lies in the sound roundoff error estimation, which is
    known to be hard and expensive~\cite{Darulova2017,Goubault2013}, and is
    largely orthogonal to the focus of this paper.
    Our error computation method can also be extended to transcendental
    functions~\cite{Darulova2011} and we plan to implement this extension in the
    future.


  We are not aware of any tool which combines rewriting and mixed-precision
  tuning and which supports both floating-point as well as fixed-point
  arithmetic.
  We experimentally compare \daisy{} with FPTuner~\cite{Chiang2017}, which is
  the only other tool for \emph{sound} (floating-point) mixed-precision tuning.
  FPTuner reduces mixed-precision tuning to an optimization problem. The
  optimization objective in FPTuner is performance, but it is up to the user to
  provide optimization constraints, e.g. in form of a maximum number of cast
  operations.
  FPTuner also supports ganging of operators where the user can specify
  constraints that limit specific operations to have the same precision, which
  can be useful for vectorization. Adding these kinds of constraints to our
  approach is straight-forward. However, this manual optimization requires user
  expertise, and we have chosen to focus on automated techniques and the
  combination of mixed-precision and rewriting. As ~\autoref{sec:mixed} shows,
  already determining the cost function without ganging is challenging.

  As FPTuner only supports floating-point arithmetic and fixed-point arithmetic
  is most useful in combination with specialized hardware, which is beyond the
  scope of this paper, we perform the experimental evaluation here for
  floating-point arithmetic only.

  We do not compare against Precimonious directly, since \daisy{} uses the same
  search algorithm internally. We would thus be merely comparing sound and
  unsound error analyses, which in our view is not meaningful.

\paragraph{Benchmarks}
  We have experimentally evaluated our approach and tool on a number of standard
  finite-precision verification benchmarks~\cite{Darulova2014,Darulova2013,Magron2015}.
  The benchmarks rigidBody, invertedPendulum and traincar are embedded controllers;
  bsplines, sine, and sqrt are examples of functions also used in the embedded domain
  (e.g. sine approximates the transcendental function whose library function is often not available,
  or expensive). The benchmarks doppler, turbine, himmilbeau and kepler are from
  the scientific computing domain.
  ~\autoref{tbl:runtimes} lists the number of arithmetic operations and
  variables for each benchmark. An asterisk ($*$) marks nonlinear benchmarks. To
  evaluate scalability, we also include four `unrolled' benchmarks (marked by
  `2x' and `3x'), where we double (or triple) the arithmetic operation count, as well
  as the number of input variables.

  Which mixed-precision assignment is possible crucially depends on the maximum allowed
  error. None of the standard benchmarks come with suitable bounds since
  the focus until now has been on uniform precision. We follow FPTuner in
  defining suitable error bounds for our benchmarks.
  For each original example program, we first compute roundoff errors for
  uniform single and double precision. Slightly rounded up, these are the error
  bounds for benchmarks denoted by $F$ and $D$ respectively. From these we
  generate error bounds which are multiples of $0.5$, $0.1$ and $0.01$ of these,
  denoted by $F_{0.5}$, $F_{0.1}$, etc. That is, we create benchmarks whose
  error bounds are half, an order of magnitude and two orders of magnitude
  smaller than in uniform precision. This corresponds to a scenario where
  uniform precision is just barely not enough and we would like to avoid the
  next higher uniform precision.

\paragraph{Experimental Setup}\label{sec:experimental-setup}
  We have performed all experiment on a Linux desktop computer with Intel Xeon
  3.30GHz and 32GB RAM. For benchmarking, we use \daisy{}'s programs generated
  in Scala (version 2.11.6) and translate FPTuner's output into Scala. This translation
  is done automatically by a tool we wrote, and does not affect the
  results, as FPTuner is platform independent. We use
  the Scalameter tool~\cite{Prokopec2012} (version 0.7) for benchmarking, which
  first warms up the JVM and detects steady-state execution \emph{after} the
  Just-In-Time compiler has run, and then benchmarks the function as it is run
  effectively in native compiled code. We use the
  \lstinline{@strictfp} annotation to ensure that the floating-point operations
  are performed exactly as specified in the program (otherwise error bounds cannot
  be guaranteed). We intentionally choose
  this setup to benchmark the mixed-precision assignments produced by \daisy{}
  and FPTuner and not compiler
  optimization effects, which are out of scope of this paper.

\paragraph{Optimization Time}

  \autoref{tbl:runtimes} compares the execution times of \daisy{} and
  FPTuner themselves. For \daisy{}, we report the times
  for mixed-precision tuning only without rewriting, which corresponds to the
  functionality that FPTuner provides, as well as the time for full
  optimization, i.e. rewriting and mixed-precision tuning.

  For each tool, we measured the running time 5 times for each benchmark variant
  separately with the bash \lstinline{time} command, recording the average real
  time. In the table, we show the aggregated time for all the variants of a
  benchmark, e.g. the total time for the $F$, $F_{0.5}$, $F_{0.1}$, ... variants
  of one benchmark together.
  These times are end-to-end, i.e. they include all phases as well as
  occasional timeouts by the backend solvers
  (Gelpia~\cite{Baranowski2016} for FPTuner and Z3 for \daisy{}).
  \daisy{} is faster than FPTuner for all benchmarks, even with
  rewriting included, and often by large factors. We suspect that this is
  due to the fact that FPTuner is solving a global optimization problem, which
  is known to be hard.

  \begin{table}
    \small
    \centering
    \renewcommand{\arraystretch}{1.0}
    \begin{tabular}{@{}lcrrr@{}}
      \toprule
      benchmark & ops - vars & FPTuner & \daisy{}-mixed & \daisy{}-full \\
      \midrule
      bspline2*  & 10 - 1  & 4m 56s  & 34s & 50s \\
      doppler* & 8 - 3 & 12m 48s & 1m 8s & 5m 4s \\
      himmilbeau*  & 15 - 2  & 9m 7s & 44s & 1m 21s  \\
      invPendulum & 7 - 4 & 3m 47s  & 32s & 45s \\
      kepler0* & 15 - 6  & 19m 17s & 43s & 1m 2s \\
      kepler1* & 24 - 4  & 1h 26m 3s & 2m 17s  & 2m 9s \\
      kepler2* & 36 - 6  & 1h 52m 38s  & 3m 36s  & 4m 22s  \\
      rigidBody1*  & 7 - 3 & 4m 45s  & 28s & 36s \\
      rigidBody2*  & 14 - 3  & 8m 0s & 43s & 1m 3s \\
      sine*  & 18 - 1  & 9m 10s  & 1m 9s & 3m 36s  \\
      sqroot*  & 14 - 1  & 4m 33s  & 40s & 1m 11s  \\
      traincar  & 28 - 14 & 17m 17s & 1m 13s  & 2m 11s  \\
      turbine1*  & 14 - 3  & 5m 15s  & 1m 22s  & 3m 56s  \\
      turbine2*  & 10 - 3  & 4m 41s  & 58s & 2m 52s  \\
      turbine3*  & 14 - 3  & 4m 23s  & 1m 21s  & 3m 44s  \\
      kepler2 (2x)  & 73 - 12 & 15h 36m 59s & 7m 57s  & 9m 40s  \\
      rigidbody2 (3x) & 44 - 9  & 58m 55s & 5m 33s  & 3m 44s  \\
      sine (3x) & 56 - 3  & 22m 40s & 8m 20s  & 13m 57s \\
      traincar (2x)  & 57 - 28 & 33m 19s & 4m 32s  & 5m 48s  \\
    \end{tabular}
    \caption{Optimization times of \daisy{} and FPTuner}
    \label{tbl:runtimes}
  \vspace{-0.5cm}
  \end{table}

  \begin{table*}
  \begin{flushleft}
    \begin{subtable}{0.45\textwidth}
      \small
      \centering
      \renewcommand{\arraystretch}{1.0}
      \begin{tabular}{@{}m{1.45cm}m{0.35cm}m{0.35cm}m{0.35cm}m{0.35cm}m{0.35cm}m{0.35cm}m{0.35cm}m{0.35cm}m{0.35cm}|c@{}}
        \toprule
        benchmark & F  & $F_{0.5}$  & $F_{0.1}$  & $F_{0.01}$ & D  & $D_{0.5}$  & $D_{0.1}$  & $D_{0.01}$ & Q & avrg\\
        \midrule
        bspline2  & \g{1.01}  & \bf{0.55}  & \g{1.00}  & \g{1.00}  & \g{1.00}  & \g{1.00}  & \g{1.00}  & \g{1.00}  & \g{1.00}  & \bf{0.95} \\
        doppler & \g{0.97}  & \bf{0.89}  & \g{0.96}  & \bf{0.95}  & \g{0.96}  & \g{1.01}  & \g{0.99}  & \g{1.00}  & \g{0.99}  & \g{0.97} \\
        himmilbeau  & \g{0.98}  & \bf{0.95}  & \g{1.02}  & \g{0.98}  & \g{1.02}  & \bf{0.61}  & \g{1.04}  & \g{1.00}  & \g{1.00}  & \g{0.96} \\
        invPend. & \g{0.98}  & \g{1.02}  & \g{1.01}  & \g{0.98}  & \g{0.99}  & \bf{0.85}  & \g{1.00}  & \g{1.01}  & \g{1.00}  & \g{0.98} \\
        kepler0 & \g{1.00}  & \x{1.07}  & \g{1.00}  & \g{1.00}  & \g{1.00}  & \bf{0.85}  & \g{0.97}  & \g{1.00}  & \g{1.00}  & \g{0.99} \\
        kepler1 & \g{1.00}  & \bf{0.90}  & \bf{0.95}  & \g{0.99}  & \g{0.98}  & \bf{0.89}  & \x{1.10}  & \g{1.00}  & \g{1.00}  & \g{0.98} \\
        kepler2 & \g{1.00}  & \g{1.02}  & \bf{0.93}  & \g{1.00}  & \g{1.00}  & \bf{0.85}  & \x{1.14}  & \g{1.00}  & \g{1.00}  & \g{0.99} \\
        rigidBody1  & \g{1.04}  & \g{1.02}  & \g{1.02}  & \g{1.00}  & \g{0.97}  & \bf{0.72}  & \bf{0.94}  & \g{1.00}  & \g{0.99}  & \g{0.97} \\
        rigidBody2  & \g{0.97}  & \bf{0.94}  & \g{1.01}  & \g{1.04}  & \g{1.01}  & \g{0.98}  & \x{1.16}  & \g{1.00}  & \g{1.00}  & \g{1.01} \\
        sine  & \g{1.01}  & \bf{0.64}  & \g{1.00}  & \g{0.99}  & \g{1.00}  & \x{1.24}  & \g{1.00}  & \g{1.00}  & \g{1.00}  & \g{0.99} \\
        sqroot  & \g{1.02}  & \bf{0.84}  & \g{0.99}  & \g{1.01}  & \g{1.00}  & \bf{0.59}  & \g{1.00}  & \g{1.00}  & \g{1.00}  & \bf{0.94} \\
        traincar & \g{0.98}  & \g{0.97}  & \g{0.98}  & \g{1.01}  & \g{1.00}  & \bf{0.61}  & \bf{0.61}  & \bf{0.87}  & \g{1.00}  & \bf{0.89} \\
        turbine1  & \g{1.00}  & \bf{0.63}  & \g{1.00}  & \g{1.01}  & \g{1.01}  & \x{1.18}  & \g{1.00}  & \g{1.00}  & \g{1.00}  & \g{0.98} \\
        turbine2  & \g{0.98}  & \bf{0.66}  & \g{0.98}  & \g{1.01}  & \g{0.98}  & \bf{0.77}  & \g{1.00}  & \g{1.00}  & \g{1.00}  & \bf{0.93} \\
        turbine3  & \g{1.00}  & \bf{0.62}  & \g{0.99}  & \g{0.99}  & \g{0.99}  & \x{1.18}  & \g{1.00}  & \g{1.01}  & \g{1.01}  & \g{0.98} \\
        kepler2 (2x)  & \g{0.98}  & \bf{0.91}  & \g{0.99}  & \g{0.99}  & \g{1.00}  & \bf{0.65}  & \x{1.13}  & \g{1.01}  & \g{1.01}  & \g{0.96}  \\
        rigidBody2(3x) & \g{0.99}  & \g{1.02}  & \g{1.00}  & \g{0.97}  & \g{0.97}  & \bf{0.74}  & \x{1.09}  & \g{1.00}  & \g{1.00}  & \g{0.97}  \\
        sine (3x) & \g{1.00}  & \bf{0.75}  & \g{1.00}  & \g{1.00}  & \g{1.00}  & \bf{0.69}  & \g{1.00}  & \g{1.00}  & \g{1.00}  & \bf{0.94}  \\
        traincar (2x)  & \x{1.13}  & \x{1.05}  & \x{1.05}  & \x{1.12}  & \x{1.14}  & \bf{0.61}  & \bf{0.87}  & \bf{0.93}  & \g{1.00}  & \g{0.99}  \\
      \end{tabular}
      \caption{\daisy{} - mixed-precision tuning only}\label{tbl:daisy-delta-only}
    \end{subtable}
    \qquad\quad
    \begin{subtable}{0.45\textwidth}
      \small
      \centering
      \renewcommand{\arraystretch}{1.0}
      \begin{tabular}{@{}m{1.45cm}m{0.35cm}m{0.35cm}m{0.35cm}m{0.35cm}m{0.35cm}m{0.35cm}m{0.35cm}m{0.35cm}m{0.35cm}|c@{}}
        \toprule
        benchmark & F  & $F_{0.5}$  & $F_{0.1}$  & $F_{0.01}$ & D  & $D_{0.5}$  & $D_{0.1}$  & $D_{0.01}$ & Q & avrg\\
        \midrule
        bspline2  & \g{0.99}  & \g{0.99}  & \g{0.99}  & \g{0.99}  & \g{0.98}  & \bf{0.90}  & \bf{0.90}  & \bf{0.90}  & \bf{0.90}  & \bf{0.95}  \\
        doppler & \g{1.02}  & \bf{0.90}  & \x{1.33}  & \x{1.32}  & \x{1.33}  & \bf{0.10}  & \g{1.01}  & \g{1.02}  & \g{1.01}  & \g{1.01}  \\
        himmilbeau  & \g{0.96}  & \g{0.98}  & \g{0.98}  & \g{0.99}  & \g{0.98}  & \g{0.99}  & \g{0.99}  & \g{0.99}  & \g{0.99}  & \g{0.98}  \\
        invPend. & \g{1.03}  & \g{0.99}  & \g{1.00}  & \g{1.03}  & \g{1.03}  & \bf{0.95}  & \bf{0.95}  & \bf{0.95}  & \g{0.96}  & \g{0.99}  \\
        kepler0 & \g{0.98}  & \g{1.02}  & \g{0.99}  & \g{1.00}  & \g{0.99}  & \g{0.97}  & \g{0.97}  & \g{0.98}  & \g{0.97}  & \g{0.99}  \\
        kepler1 & \g{0.99}  & \bf{0.87}  & \bf{0.87}  & \bf{0.87}  & \bf{0.87}  & \g{1.00}  & \g{1.00}  & \g{1.00}  & \g{1.00}  & \bf{0.94}  \\
        kepler2 & \g{0.96}  & \bf{0.92}  & \bf{0.92}  & \bf{0.95}  & \bf{0.93}  & \g{0.99}  & \g{0.99}  & \g{0.99}  & \g{0.99}  & \g{0.96}  \\
        rigidBody1  & \g{0.99}  & \g{0.99}  & \g{1.01}  & \g{1.01}  & \g{0.99}  & \bf{0.93}  & \bf{0.93}  & \bf{0.93}  & \bf{0.93}  & \g{0.97}  \\
        rigidBody2  & \g{0.96}  & \bf{0.88}  & \bf{0.89}  & \bf{0.90}  & \bf{0.90}  & \g{0.98}  & \g{0.98}  & \g{0.98}  & \g{0.98}  & \bf{0.94}  \\
        sine  & \g{0.99}  & \x{1.08}  & \x{1.08}  & \x{1.08}  & \x{1.09}  & \g{0.98}  & \g{0.98}  & \g{0.98}  & \g{0.98}  & \g{1.03}  \\
        sqroot  & \g{0.96}  & \bf{0.95}  & \bf{0.93}  & \bf{0.92}  & \bf{0.93}  & \g{0.97}  & \g{0.97}  & \g{0.98}  & \g{0.98}  & \bf{0.95}  \\
        traincar & \bf{0.89}  & \bf{0.85}  & \bf{0.89}  & \bf{0.89}  & \bf{0.91}  & \bf{0.07}  & \g{1.02}  & \g{1.02}  & \g{1.02}  & \bf{0.84}  \\
        turbine1  & \g{1.02}  & \g{0.97}  & \g{0.97}  & \g{0.96}  & \g{0.96}  & \x{1.06}  & \x{1.06}  & \x{1.06}  & \x{1.06}  & \g{1.01}  \\
        turbine2  & \bf{0.94}  & \bf{0.94}  & \bf{0.94}  & \bf{0.94}  & \bf{0.95}  & \g{0.99}  & \g{0.99}  & \g{0.98}  & \g{0.99}  & \g{0.96}  \\
        turbine3  & \g{1.01}  & \g{0.96}  & \g{0.97}  & \bf{0.95}  & \bf{0.95}  & \x{1.07}  & \x{1.07}  & \x{1.07}  & \x{1.07}  & \g{1.01}  \\
        kepler2 (2x)  & \bf{0.89}  & \bf{0.78}  & \bf{0.86}  & \bf{0.87}  & \bf{0.88}  & \g{0.96}  & \g{0.98}  & \g{1.00}  & \g{0.96}  & \bf{0.91}  \\
        rigidBody2(3x) & \g{1.04}  & \g{0.96}  & \bf{0.94}  & \bf{0.94}  & \g{0.98}  & \g{1.00}  & \g{0.99}  & \g{1.00}  & \g{1.01}  & \g{0.99}  \\
        sine (3x) & \bf{0.93}  & \x{1.05}  & \bf{0.91}  & \bf{0.90}  & \bf{0.90}  & \g{0.99}  & \g{0.99}  & \g{1.00}  & \g{0.98}  & \g{0.96}  \\
        traincar (2x)  & \g{0.98}  & \bf{0.90}  & \g{0.98}  & \g{0.99}  & \g{1.01}  & \bf{0.09}  & \g{1.03}  & \g{1.02}  & \g{1.02}  & \bf{0.89}  \\
      \end{tabular}
      \caption{\daisy{} - rewriting only}\label{tbl:daisy-rewriting-only}
    \end{subtable}

    \begin{subtable}{0.45\textwidth}
      \small
      \centering
      \renewcommand{\arraystretch}{1.0}
      \begin{tabular}{@{}m{1.45cm}m{0.35cm}m{0.35cm}m{0.35cm}m{0.35cm}m{0.35cm}m{0.35cm}m{0.35cm}m{0.35cm}m{0.35cm}|c@{}}
        \toprule
        benchmark & F  & $F_{0.5}$  & $F_{0.1}$  & $F_{0.01}$ & D  & $D_{0.5}$  & $D_{0.1}$  & $D_{0.01}$ & Q & avrg\\
        \midrule
        bspline2  & \g{1.04}  & \bf{0.50}  & \g{0.99}  & \g{0.97}  & \g{0.98}  & \bf{0.77}  & \bf{0.90}  & \bf{0.91}  & \bf{0.91}  & \bf{0.88}  \\
        doppler & \g{1.03}  & \bf{0.87}  & \g{0.98}  & \x{1.31}  & \x{1.31}  & \bf{0.10}  & \x{1.18}  & \g{1.02}  & \g{1.01}  & \g{0.98}  \\
        himmilbeau  & \g{0.99}  & \g{0.96}  & \g{1.01}  & \g{1.00}  & \g{0.98}  & \bf{0.60}  & \x{1.07}  & \g{0.99}  & \g{0.99}  & \bf{0.95}  \\
        invPend. & \g{0.99}  & \g{0.99}  & \g{0.99}  & \g{0.97}  & \g{0.97}  & \bf{0.68}  & \g{0.97}  & \bf{0.95}  & \bf{0.95}  & \bf{0.94}  \\
        kepler0 & \bf{0.94}  & \x{1.05}  & \g{0.99}  & \g{0.96}  & \g{0.96}  & \bf{0.55}  & \g{1.00}  & \g{0.98}  & \g{0.98}  & \bf{0.93}  \\
        kepler1 & \g{0.98}  & \bf{0.91}  & \bf{0.87}  & \bf{0.90}  & \bf{0.87}  & \bf{0.89}  & \g{0.98}  & \g{1.00}  & \g{1.00}  & \bf{0.93}  \\
        kepler2 & \g{0.96}  & \g{1.04}  & \bf{0.94}  & \bf{0.93}  & \bf{0.93}  & \bf{0.55}  & \x{1.08}  & \g{0.98}  & \g{0.98}  & \bf{0.93}  \\
        rigidBody1  & \g{1.00}  & \g{1.02}  & \g{0.97}  & \g{0.96}  & \g{0.99}  & \bf{0.57}  & \g{1.04}  & \bf{0.93}  & \bf{0.94}  & \bf{0.94}  \\
        rigidBody2  & \g{0.97}  & \g{0.98}  & \bf{0.90}  & \bf{0.88}  & \bf{0.88}  & \bf{0.61}  &\x{1.16}  & \g{0.98}  & \g{0.98}  & \bf{0.93}  \\
        sine  & \g{0.98}  & \bf{0.65}  & \x{1.09}  & \x{1.10}  & \x{1.09}  & \x{1.26}  & \g{0.99}  & \g{0.98}  & \g{0.98}  & \g{1.01}  \\
        sqroot  & \g{0.97}  & \bf{0.87}  & \bf{0.95}  & \bf{0.92}  & \bf{0.93}  & \bf{0.64}  & \g{1.02}  & \g{0.97}  & \g{0.97}  & \bf{0.92}  \\
        traincar & \bf{0.88}  & \bf{0.85}  & \bf{0.86}  & \bf{0.89}  & \bf{0.91}  & \bf{0.07}  & \bf{0.66}  & \bf{0.90}  & \g{1.02}  & \bf{0.78}  \\
        turbine1  & \g{0.99}  & \bf{0.68}  & \g{0.96}  & \g{0.96}  & \g{0.96}  & \x{1.17}  & \x{1.06}  & \x{1.06}  & \x{1.06}  & \g{0.99}  \\
        turbine2  & \g{0.98}  & \bf{0.67}  & \bf{0.94}  & \bf{0.95}  & \bf{0.95}  & \bf{0.74}  & \g{0.99}  & \g{0.99}  & \g{1.00}  & \bf{0.91}  \\
        turbine3  & \g{1.00}  & \bf{0.70}  & \g{0.96}  & \g{0.96}  & \bf{0.95}  & \x{1.16}  & \x{1.07}  & \x{1.07}  & \x{1.07}  & \g{0.99}  \\
        kepler2 (2x)  & \bf{0.86}  & \bf{0.79}  & \bf{0.86}  & \g{0.96}  & \bf{0.89}  & \bf{0.80}  & \x{1.05}  & \g{1.00}  & \g{0.97}  & \bf{0.91}  \\
        rigidBody2(3x) & \g{1.00}  & \g{1.02}  & \bf{0.94}  & \bf{0.95}  & \g{0.96}  & \bf{0.55}  & \g{0.98}  & \g{1.01}  & \g{0.99}  & \bf{0.93}  \\
        sine (3x) & \bf{0.93}  & \bf{0.61}  & \bf{0.95}  & \bf{0.90}  & \bf{0.89}  & \bf{0.46}  & \g{1.03}  & \g{0.99}  & \g{0.99}  & \bf{0.86}  \\
        traincar (2x)  & \g{0.96}  & \bf{0.91}  & \bf{0.93}  & \g{0.97}  & \g{1.00}  & \bf{0.08}  & \bf{0.91}  & \g{0.99}  & \g{1.03}  & \bf{0.87}  \\
      \end{tabular}
      \caption{\daisy{} - full optimization}
    \end{subtable}
    \qquad\quad
    \begin{subtable}{0.45\textwidth}
      \small
      \centering
      \renewcommand{\arraystretch}{1.0}
      \begin{tabular}{@{}m{1.45cm}m{0.35cm}m{0.35cm}m{0.35cm}m{0.35cm}m{0.35cm}m{0.35cm}m{0.35cm}m{0.35cm}m{0.35cm}|c@{}}
        \toprule
        benchmark & F  & $F_{0.5}$  & $F_{0.1}$  & $F_{0.01}$ & D  & $D_{0.5}$  & $D_{0.1}$  & $D_{0.01}$ & Q & avrg\\
        \midrule
        bspline2  & \g{1.00}  & \bf{0.53}  & \g{1.01}  & \g{1.00}  & \g{1.00}  & \bf{0.65}  & \x{1.21}  & \x{1.21}  & \x{1.22}  & \g{0.98}  \\
        doppler & \g{0.99}  & \bf{0.87}  & \g{1.03}  & \g{1.01}  & \g{1.00}  & \x{0.07}  & \x{1.10}  & \x{1.57}  & \x{1.62}  & \g{1.03}  \\
        himmilbeau  & \g{0.98}  & \bf{0.88}  & \x{1.22}  & \g{0.99}  & \g{1.01}  & \x{0.19}  & \bf{0.92}  & \x{1.06}  & \x{1.05}  & \bf{0.92}  \\
        invPend. & \x{1.05}  & \g{1.01}  & \g{1.04}  & \g{0.98}  & \g{0.99}  & \bf{0.61}  & \g{0.98}  & \x{1.07}  & \x{1.05}  & \g{0.98}  \\
        kepler0 & \x{1.06}  & \x{1.06}  & \x{1.07}  & \g{1.01}  & \g{1.01}  & \bf{0.49}  & \g{1.01}  & \x{1.10}  & \x{1.10}  & \g{0.99}  \\
        kepler1 & \g{1.03}  & \g{1.02}  & \x{1.22}  & \g{0.99}  & \x{5.46}  & \bf{0.57}  & \g{0.99}  & \x{1.07}  & \x{1.07}  & \x{1.49}  \\
        kepler2 & \g{0.98}  & \x{1.05}  & \x{1.11}  & \x{1.24}  & \x{3.03}  & \bf{0.64}  & \bf{0.88}  & \g{1.04}  & \g{1.04}  & \x{1.22}  \\
        rigidBody1  & \x{1.08}  & \g{1.03}  & \g{0.99}  & \g{0.96}  & \x{4.66}  & \bf{0.61}  & \g{1.03}  & \x{1.21}  & \x{1.21}  & \x{1.42}  \\
        rigidBody2  & \x{1.08}  & \bf{0.95}  & \x{1.18}  & \g{1.02}  & \x{5.52}  & \bf{0.56}  & \bf{0.57}  & \bf{0.89}  & \x{1.13}  & \x{1.43}  \\
        sine  & \g{1.01}  & \bf{0.43}  & \bf{0.86}  & \x{0.93}  & \g{1.00}  & \bf{0.20}  & \bf{0.48}  & \bf{0.68}  & \g{1.03}  & \bf{0.74}  \\
        sqroot  & \x{1.14}  & \bf{0.83}  & \x{1.22}  & \x{1.12}  & \x{4.86}  & \bf{0.40}  & \bf{0.68}  & \bf{0.94}  & \x{1.24}  & \x{1.38}  \\
        traincar & \g{1.01}  & \bf{0.90}  & \bf{0.94}  & \bf{0.92}  & \x{4.55}  & \bf{0.36}  & \bf{0.37}  & \bf{0.37}  & \g{1.04}  & \x{1.16}  \\
        turbine1  & \g{1.00}  & \bf{0.56}  & \bf{0.88}  & \bf{0.95}  & \g{1.00}  & \bf{0.07}  & \bf{0.86}  & \bf{0.92}  & \x{1.18}  & \bf{0.82}  \\
        turbine2  & \g{1.00}  & \bf{0.61}  & \bf{0.79}  & \g{0.98}  & \g{1.00}  & \bf{0.09}  & \bf{0.92}  & \x{1.13}  & \x{1.12}  & \bf{0.85}  \\
        turbine3  & \g{1.01}  & \bf{0.58}  & \g{0.96}  & \bf{0.92}  & \g{1.00}  & \bf{0.07}  & \bf{0.68}  & \bf{0.91}  & \x{1.18}  & \bf{0.81}  \\
        kepler2 (2x)  & \g{0.99}  & \g{0.97}  & \x{1.13}  & \x{1.09}  & \x{2.91}  & \bf{0.55}  & \bf{0.86}  & \g{crash} & \g{1.03}  & \x{1.19}  \\
        rigidBody2(3x) & \g{1.03}  & \g{1.03}  & \x{1.13}  & \x{1.08}  & \x{4.97}  & \bf{0.45}  & \bf{0.61}  & \bf{0.83}  & \x{1.13} & \x{1.36}  \\
        sine (3x) & \g{0.99}  & \bf{0.53}  & \bf{0.85}  & \g{1.02}  & \x{2.20}  & \bf{0.19}  & \bf{0.45}  & \bf{0.69}  & \g{1.02}  & \bf{0.88}  \\
        traincar (2x)  & \x{1.11}  & \g{0.96}  & \g{0.99}  & \g{0.99}  & \x{5.26}  & \bf{0.50}  & \bf{0.51}  & \bf{0.51}  & \g{1.03}  & \x{1.32}  \\
      \end{tabular}
      \caption{FPTuner}
    \end{subtable}
    \caption{Relative performance improvements for \daisy{} and FPTuner}\label{tbl:end-to-end}
    \vspace{-0.75cm}
    \end{flushleft}
  \end{table*}

\paragraph{Performance Improvements}

  To evaluate the effectiveness of \daisy{} we have performed end-to-end
  performance experiments. For this, we benchmark each generated optimized program five times with
  Scalameter and use the average running time. Each run of a program consists of
  100 000 executions on random inputs.
  Then, for each mixed-precision variant we compare its running time
  against the corresponding uniform-precision program and report the relative
  improvement (i.e. mixed-precision running time/ uniform running time).
  Corresponding here means the smallest uniform precision which would satisfy
  the error bound, e.g. for the $F_{0.5}$ benchmark, the smallest uniform precision
  satisfying this bound is double precision.

  \autoref{tbl:end-to-end} shows the relative running time improvements for
  \daisy{} with only mixed-precision tuning, with only rewriting and with both
  rewriting and mixed-precision tuning enabled as well as for FPTuner.
  We show the average speedups over all variants of a benchmark (row).
  Bold values mark performance improvements above 5\% and underlined values those
  with over 5\% slowdowns.

  Overall, we can see that the biggest speedups occur for the
  $F_{0.5}$ and $D_{0.5}$ benchmarks, as expected.
  The very low values (e.g. 0.07 in ~\autoref{tbl:daisy-rewriting-only} for
  train4-st8-$D_{0.5}$) result from rewriting reducing the roundoff error sufficiently
  for uniform precision being enough (the baseline comparison running time is however the
  higher uniform precision).
  In the case of FPTuner, the very low and very high values are caused by
  different characteristics of the error analyses. As a consequence, FPTuner is
  able to compute smaller roundoff errors than \daisy{} for some benchmarks (and
  assigning uniform double precision instead of mixed), while for others it
  computes bigger roundoff errors, where it cannot show that double precision is
  sufficient, whereas \daisy{} can.

  Comparing FPTuner with \daisy{}'s mixed-precision tuning only
  (\autoref{tbl:daisy-delta-only}), we observe that \daisy{} is more
  conservative, both with performance improvements but also with slowdowns,
  increasing the execution time only rarely.
  Considering \daisy{}'s significantly better optimization time,
  \daisy{} provides an interesting tradeoff between mixed-precision tuning and
  efficiency.

  Comparing the speedups obtained by mixed-precision tuning and rewriting, we
  note that both techniques are successfull in improving the performance, though
  the effect of rewriting is modest.
  When the two techniques are combined, we obtain the biggest performance
  improvements, which are furthermore more than just the sum of both.
  Note that rewriting could also be
  combined with (i.e. run before) FPTuner's mixed-precision tuning.
\section{Related Work}\label{sec:related}


  \paragraph{Rewriting}
    An alternative approach to rewriting was presented by
    \citeauthor{Damouche2015}~\cite{Damouche2015} which relies on a greedy
    search and an abstract domain which represents possible expression
    rewritings together with a static error analysis similar to ours.
    The tool Herbie~\cite{Panchekha2015} performs a greedy
    hill-climbing search guided by a dynamic error evaluation function, and as
    such cannot provide sound error bounds. It is geared more towards correcting
    catastrophic cancellations, by employing an `error localization' function
    which pin-points an operation which commits a particularly large roundoff
    error and then targets the rewriting rules at that part of the expression.
    %
    It would be interesting in the future to compare these different search techniques
    for rewriting.

  \paragraph{Mixed-precision Tuning in HPC}
  Mixed-precision is especially important in HPC applications, because
  floating-point arithmetic is widely used.
  \citeauthor{Lam2013b} introduced a binary intrumentation tool together with a
  breadth-first search algorithm to help programmers search for suitable
  mixed-precision programs~\cite{Lam2013b}. This work was later extended to
  perform a sensistivity analysis~\cite{Lam2016} based on a more fine-grained
  approach. The Precimonious
  project~\cite{Rubio-Gonzalez2013,Rubio-Gonzales2016}, whose delta-debugging
  algorithm we adapt, targets HPC kernels and library functions and performs
  automated mixed-precision tuning. These projects have in common that the roundoff
  error verification is performed dynamically on a limited number of
  inputs and thus does not provide guarantees. In contrast,
  our technique produces sound results, but is targeted at smaller programs and
  kernels which can be verified statically.

  \paragraph{Autotuning} Another way to improve the performance of (numerical)
  computations is autotuning, which performs low-level transformations of the
  program in order to find one which empirically executes most efficiently.
  Traditionally, the approaches have been semantics
  preserving~\cite{Puschel2004,Vuduc2004}, but recently also non-semantics
  preserving ones have been proposed in the space of approximate
  computing~\cite{Schkufza2014}. These techniques represent another avenue for
  improving performance, but do not optimize mixed-precision.

  \paragraph{Bitlength Optimization in Embedded Systems}
  In the space of embedded systems, much of the attention so far
  has focused on fixed-point arithmetic and the
  optimization of bitlengths, which can be viewed as selecting data types.
  A variety of static and dynamic approaches have been applied.
  For instance, \citeauthor{Gaffar2004} considers both fixed-point and
  floating-point programs and uses automatic differentiation for a sensitivity
  analysis~\cite{Gaffar2004}. \citeauthor{Mallik2007} present an optimal
  bit-width allocation for two variables and a greedy heuristic for more
  variables, and rely on dynamic error evaluation~\cite{Mallik2007}. Unlike our
  approach, these two techniques cannot provide sound error bounds.
  Sound techniques have also been applied for both the range and the error
  analysis for bitwidth optimization, for instance
  in~\cite{Lee2006,Osborne2007,Kinsman2009,Pang2011} and \citeauthor{Lee2006}
  provide a nice overview of static and dynamic techniques~\cite{Lee2006}. For
  optimization, \citeauthor{Lee2006} have used simulated annealing as the search
  technique~\cite{Lee2006}. A detailed comparison of delta-debugging and e.g.
  simulated annealing would be very interesting in the future. We note that our
  technique is general in that it is applicable to both floating-point as well
  as fixed-point arithmetic, and the first to combine bitwidth optimization for
  performance with rewriting.

  \paragraph{Finite-precision Verification}

  There has been considerable interest in static and sound numerical error
  estimation for finite-precision programs with several tools having been
  developed: Rosa~\cite{Darulova2014}, Fluctuat~\cite{Goubault2011},
  FPTaylor~\cite{Solovyev2015} (which FPTuner is based on) and \\
  Real2Float~\cite{Magron2015}. The accuracies of these
  tools are mostly comparable~\cite{Darulova2017}, so that any
  of the underlying techniques could be used in our approach for the static
  error function.
  More broadly related are abstract interpretation-based static analyses, which
  are sound wrt. floating-point
  arithmetic~\cite{Blanchet2003,Chen2008,Jeannet2009}. These techniques can
  prove the absence of runtime errors, such as division-by-zero, but cannot
  quantify roundoff errors.
  Floating-point arithmetic has also been formalized in theorem provers such as
  Coq~\cite{Boldo2011}~\cite{Daumas_FP} and HOL Light~\cite{Jacobsen2015}, and
  entire numerical programs have been proven correct and accurate within
  these~\cite{Boldo2013,Ramananandro2016}. Most of these verification efforts
  are to a large part manual, and do not perform mixed-precision tuning.
  FPTaylor uses HOL Light and Real2Float Coq to generate certificates of correctness of
  the error bounds it computes. We believe that this facility could be
  extended to the mixed-precision case -- this, however, would come after the
  tuning step, and hence these efforts are largely orthogonal.

  Floating-point arithmetic has also been formalized in an SMT-lib
  ~\cite{Rummer2010} theory and SMT solvers exist which include floating-point
  decision procedures~\cite{De-Moura2008,Brain2013}. These are, however, not
  suitable for roundoff error quantification, as a combination with the theory
  of reals would be necessary which does not exist today.

\section{Conclusion}


  We have presented a fully automated technique which combines rewriting and
  sound mixed-precision tuning for improving the performance of arithmetic
  kernels. While each of the two parts is successful by itself, we have
  empirically demonstrated that their careful combination is more than just the sum of
  the parts. Furthermore, our mixed-precision tuning algorithm presents an
  interesting tradeoff as compared to state-of-the-art between efficiency of the
  algorithm and performance improvements generated.





\bibliographystyle{ACM-Reference-Format}
\bibliography{main}


\begin{thebibliography}{00}


\ifx \showCODEN    \undefined \def \showCODEN     #1{\unskip}     \fi
\ifx \showDOI      \undefined \def \showDOI       #1{{\tt DOI:}\penalty0{#1}\ }
  \fi
\ifx \showISBNx    \undefined \def \showISBNx     #1{\unskip}     \fi
\ifx \showISBNxiii \undefined \def \showISBNxiii  #1{\unskip}     \fi
\ifx \showISSN     \undefined \def \showISSN      #1{\unskip}     \fi
\ifx \showLCCN     \undefined \def \showLCCN      #1{\unskip}     \fi
\ifx \shownote     \undefined \def \shownote      #1{#1}          \fi
\ifx \showarticletitle \undefined \def \showarticletitle #1{#1}   \fi
\ifx \showURL      \undefined \def \showURL       #1{#1}          \fi
\providecommand\bibfield[2]{#2}
\providecommand\bibinfo[2]{#2}
\providecommand\natexlab[1]{#1}
\providecommand\showeprint[2][]{arXiv:#2}

\bibitem[\protect\citeauthoryear{Anta, Majumdar, Saha, and Tabuada}{Anta
  et~al\mbox{.}}{2010}]%
        {Anta2010}
\bibfield{author}{\bibinfo{person}{A.~ Anta}, \bibinfo{person}{R.~ Majumdar},
  \bibinfo{person}{I.~ Saha}, {and} \bibinfo{person}{P.~ Tabuada}.}
  \bibinfo{year}{2010}\natexlab{}.
\newblock \showarticletitle{Automatic {V}erification of {C}ontrol {S}ystem
  {I}mplementations}. In \bibinfo{booktitle}{{\em EMSOFT}}.
\newblock


\bibitem[\protect\citeauthoryear{Bailey, Hida, Li, and Thompson}{Bailey
  et~al\mbox{.}}{2015}]%
        {Bailey2015}
\bibfield{author}{\bibinfo{person}{D.~H.~ Bailey}, \bibinfo{person}{Y.~ Hida},
  \bibinfo{person}{X.~S.~ Li}, {and} \bibinfo{person}{B.~ Thompson}.}
  \bibinfo{year}{2015}\natexlab{}.
\newblock \bibinfo{booktitle}{{\em {C++/Fortran-90 double-double and
  quad-double package}}}.
\newblock \bibinfo{type}{{T}echnical {R}eport}.
\newblock
\showURL{%
\url{http://crd-legacy.lbl.gov/~dhbailey/mpdist/}}


\bibitem[\protect\citeauthoryear{Baranowski and Briggs}{Baranowski and
  Briggs}{2016}]%
        {Baranowski2016}
\bibfield{author}{\bibinfo{person}{M.~S.~ Baranowski} {and}
  \bibinfo{person}{I.~ Briggs}.} \bibinfo{year}{2016}\natexlab{}.
\newblock \bibinfo{title}{{Global Extrema Locator Parallelization for Interval
  Arithmetic (Gelpia)}}.
\newblock \bibinfo{howpublished}{\url{https://github.com/soarlab/gelpia}}.
  (\bibinfo{year}{2016}).
\newblock


\bibitem[\protect\citeauthoryear{Blanchet, Cousot, Cousot, Feret, Mauborgne,
  Min{\'e}, Monniaux, and Rival}{Blanchet et~al\mbox{.}}{2003}]%
        {Blanchet2003}
\bibfield{author}{\bibinfo{person}{B.~ Blanchet}, \bibinfo{person}{P.~ Cousot},
  \bibinfo{person}{R.~ Cousot}, \bibinfo{person}{J.~ Feret},
  \bibinfo{person}{L.~ Mauborgne}, \bibinfo{person}{A.~ Min{\'e}},
  \bibinfo{person}{D.~ Monniaux}, {and} \bibinfo{person}{X.~ Rival}.}
  \bibinfo{year}{2003}\natexlab{}.
\newblock \showarticletitle{{A Static Analyzer for Large Safety-Critical
  Software}}. In \bibinfo{booktitle}{{\em PLDI}}.
\newblock


\bibitem[\protect\citeauthoryear{Boldo, Cl{\'e}ment, Filli{\^a}tre, Mayero,
  Melquiond, and Weis}{Boldo et~al\mbox{.}}{2013}]%
        {Boldo2013}
\bibfield{author}{\bibinfo{person}{S.~ Boldo}, \bibinfo{person}{F.~
  Cl{\'e}ment}, \bibinfo{person}{J.-C.~ Filli{\^a}tre}, \bibinfo{person}{M.~
  Mayero}, \bibinfo{person}{G.~ Melquiond}, {and} \bibinfo{person}{P.~ Weis}.}
  \bibinfo{year}{2013}\natexlab{}.
\newblock \showarticletitle{{Wave Equation Numerical Resolution: A
  Comprehensive Mechanized Proof of a C Program}}.
\newblock \bibinfo{journal}{{\em Journal of Automated Reasoning\/}}
  \bibinfo{volume}{50}, \bibinfo{number}{4} (\bibinfo{year}{2013}),
  \bibinfo{pages}{423--456}.
\newblock


\bibitem[\protect\citeauthoryear{Boldo and Melquiond}{Boldo and
  Melquiond}{2011}]%
        {Boldo2011}
\bibfield{author}{\bibinfo{person}{S.~ Boldo} {and} \bibinfo{person}{G.~
  Melquiond}.} \bibinfo{year}{2011}\natexlab{}.
\newblock \showarticletitle{{Flocq: A Unified Library for Proving
  Floating-Point Algorithms in Coq}}. In \bibinfo{booktitle}{{\em ARITH}}.
\newblock


\bibitem[\protect\citeauthoryear{Brain, D'Silva, Griggio, Haller, and
  Kroening}{Brain et~al\mbox{.}}{2013}]%
        {Brain2013}
\bibfield{author}{\bibinfo{person}{M.~ Brain}, \bibinfo{person}{V.~ D'Silva},
  \bibinfo{person}{A.~ Griggio}, \bibinfo{person}{L.~ Haller}, {and}
  \bibinfo{person}{D.~ Kroening}.} \bibinfo{year}{2013}\natexlab{}.
\newblock \showarticletitle{{Deciding floating-point logic with abstract
  conflict driven clause learning}}.
\newblock \bibinfo{journal}{{\em Formal Methods in System Design\/}}
  \bibinfo{volume}{45}, \bibinfo{number}{2} (\bibinfo{date}{Dec.}
  \bibinfo{year}{2013}), \bibinfo{pages}{213--245}.
\newblock


\bibitem[\protect\citeauthoryear{Chen, Min{\'e}, and Cousot}{Chen
  et~al\mbox{.}}{2008}]%
        {Chen2008}
\bibfield{author}{\bibinfo{person}{L.~ Chen}, \bibinfo{person}{A.~ Min{\'e}},
  {and} \bibinfo{person}{P.~ Cousot}.} \bibinfo{year}{2008}\natexlab{}.
\newblock \showarticletitle{A {S}ound {F}loating-{P}oint {P}olyhedra {A}bstract
  {D}omain}. In \bibinfo{booktitle}{{\em APLAS}}.
\newblock


\bibitem[\protect\citeauthoryear{Chiang, Gopalakrishnan, Rakamaric, Briggs,
  Baranowski, and Solovyev}{Chiang et~al\mbox{.}}{2017}]%
        {Chiang2017}
\bibfield{author}{\bibinfo{person}{W.-F.~ Chiang}, \bibinfo{person}{G.~
  Gopalakrishnan}, \bibinfo{person}{Z.~ Rakamaric}, \bibinfo{person}{I.~
  Briggs}, \bibinfo{person}{M.~S.~ Baranowski}, {and} \bibinfo{person}{A.~
  Solovyev}.} \bibinfo{year}{2017}\natexlab{}.
\newblock \showarticletitle{{Rigorous Floating-point Mixed Precision Tuning}}.
  In \bibinfo{booktitle}{{\em POPL}}.
\newblock


\bibitem[\protect\citeauthoryear{Damouche, Martel, and Chapoutot}{Damouche
  et~al\mbox{.}}{2015}]%
        {Damouche2015}
\bibfield{author}{\bibinfo{person}{N.~ Damouche}, \bibinfo{person}{M.~ Martel},
  {and} \bibinfo{person}{A.~ Chapoutot}.} \bibinfo{year}{2015}\natexlab{}.
\newblock \showarticletitle{{Intra-procedural Optimization of the Numerical
  Accuracy of Programs}}. In \bibinfo{booktitle}{{\em FMICS}}.
\newblock


\bibitem[\protect\citeauthoryear{Darulova and Kuncak}{Darulova and
  Kuncak}{2011}]%
        {Darulova2011}
\bibfield{author}{\bibinfo{person}{E.~ Darulova} {and} \bibinfo{person}{V.~
  Kuncak}.} \bibinfo{year}{2011}\natexlab{}.
\newblock \showarticletitle{{Trustworthy Numerical Computation in Scala}}. In
  \bibinfo{booktitle}{{\em OOPSLA}}.
\newblock


\bibitem[\protect\citeauthoryear{Darulova and Kuncak}{Darulova and
  Kuncak}{2014}]%
        {Darulova2014}
\bibfield{author}{\bibinfo{person}{E.~ Darulova} {and} \bibinfo{person}{V.~
  Kuncak}.} \bibinfo{year}{2014}\natexlab{}.
\newblock \showarticletitle{{Sound Compilation of Reals}}. In
  \bibinfo{booktitle}{{\em POPL}}.
\newblock


\bibitem[\protect\citeauthoryear{Darulova and Kuncak}{Darulova and
  Kuncak}{2017}]%
        {Darulova2017}
\bibfield{author}{\bibinfo{person}{E.~ Darulova} {and} \bibinfo{person}{V.~
  Kuncak}.} \bibinfo{year}{2017}\natexlab{}.
\newblock \showarticletitle{Towards a Compiler for Reals}.
\newblock \bibinfo{journal}{{\em ACM TOPLAS\/}} \bibinfo{volume}{39},
  \bibinfo{number}{2} (\bibinfo{year}{2017}).
\newblock


\bibitem[\protect\citeauthoryear{Darulova, Kuncak, Majumdar, and Saha}{Darulova
  et~al\mbox{.}}{2013}]%
        {Darulova2013}
\bibfield{author}{\bibinfo{person}{E.~ Darulova}, \bibinfo{person}{V.~ Kuncak},
  \bibinfo{person}{R.~ Majumdar}, {and} \bibinfo{person}{I.~ Saha}.}
  \bibinfo{year}{2013}\natexlab{}.
\newblock \showarticletitle{{Synthesis of Fixed-point Programs}}. In
  \bibinfo{booktitle}{{\em EMSOFT}}.
\newblock


\bibitem[\protect\citeauthoryear{Daumas, Rideau, and Th{\'e}ry}{Daumas
  et~al\mbox{.}}{2001}]%
        {Daumas_FP}
\bibfield{author}{\bibinfo{person}{M.~ Daumas}, \bibinfo{person}{L.~ Rideau},
  {and} \bibinfo{person}{L.~ Th{\'e}ry}.} \bibinfo{year}{2001}\natexlab{}.
\newblock \showarticletitle{{A Generic Library for Floating-Point Numbers and
  Its Application to Exact Computing}}. In \bibinfo{booktitle}{{\em TPHOLs}}.
\newblock


\bibitem[\protect\citeauthoryear{de~Figueiredo and Stolfi}{de~Figueiredo and
  Stolfi}{2004}]%
        {Figueiredo2004}
\bibfield{author}{\bibinfo{person}{L.~H.~ de Figueiredo} {and}
  \bibinfo{person}{J.~ Stolfi}.} \bibinfo{year}{2004}\natexlab{}.
\newblock \showarticletitle{Affine {A}rithmetic: {C}oncepts and
  {A}pplications}.
\newblock \bibinfo{journal}{{\em Numerical Algorithms\/}} \bibinfo{volume}{37},
  \bibinfo{number}{1-4} (\bibinfo{year}{2004}).
\newblock


\bibitem[\protect\citeauthoryear{De~Moura and Bj{\o}rner}{De~Moura and
  Bj{\o}rner}{2008}]%
        {De-Moura2008}
\bibfield{author}{\bibinfo{person}{L.~ De~Moura} {and} \bibinfo{person}{N.~
  Bj{\o}rner}.} \bibinfo{year}{2008}\natexlab{}.
\newblock \showarticletitle{Z3: an efficient {SMT} solver}. In
  \bibinfo{booktitle}{{\em TACAS}}.
\newblock


\bibitem[\protect\citeauthoryear{Gaffar, Mencer, Luk, and Cheung}{Gaffar
  et~al\mbox{.}}{2004}]%
        {Gaffar2004}
\bibfield{author}{\bibinfo{person}{A.~A.~ Gaffar}, \bibinfo{person}{O.~
  Mencer}, \bibinfo{person}{W.~ Luk}, {and} \bibinfo{person}{P.~Y.~K.~
  Cheung}.} \bibinfo{year}{2004}\natexlab{}.
\newblock \showarticletitle{{Unifying Bit-Width Optimisation for Fixed-Point
  and Floating-Point Designs.}}
\newblock \bibinfo{journal}{{\em FCCM\/}} (\bibinfo{year}{2004}).
\newblock


\bibitem[\protect\citeauthoryear{Goubault and Putot}{Goubault and
  Putot}{2011}]%
        {Goubault2011}
\bibfield{author}{\bibinfo{person}{E.~ Goubault} {and} \bibinfo{person}{S.~
  Putot}.} \bibinfo{year}{2011}\natexlab{}.
\newblock \showarticletitle{{Static Analysis of Finite Precision
  Computations}}. In \bibinfo{booktitle}{{\em VMCAI}}.
\newblock


\bibitem[\protect\citeauthoryear{Goubault and Putot}{Goubault and
  Putot}{2013}]%
        {Goubault2013}
\bibfield{author}{\bibinfo{person}{E.~ Goubault} {and} \bibinfo{person}{S.~
  Putot}.} \bibinfo{year}{2013}\natexlab{}.
\newblock \showarticletitle{{Robustness Analysis of Finite Precision
  Implementations}}. In \bibinfo{booktitle}{{\em APLAS}}.
\newblock


\bibitem[\protect\citeauthoryear{ISO/IEC}{ISO/IEC}{2008}]%
        {ISOIEC2008}
\bibfield{author}{\bibinfo{person}{ISO/IEC}.} \bibinfo{year}{2008}\natexlab{}.
\newblock \bibinfo{booktitle}{{\em {Programming languages --- C --- Extensions
  to support embedded processors}}}.
\newblock \bibinfo{type}{{T}echnical {R}eport} ISO/IEC TR 18037.
\newblock


\bibitem[\protect\citeauthoryear{Jacobsen, Solovyev, and
  Gopalakrishnan}{Jacobsen et~al\mbox{.}}{2015}]%
        {Jacobsen2015}
\bibfield{author}{\bibinfo{person}{C.~ Jacobsen}, \bibinfo{person}{A.~
  Solovyev}, {and} \bibinfo{person}{G.~ Gopalakrishnan}.}
  \bibinfo{year}{2015}\natexlab{}.
\newblock \showarticletitle{{A Parameterized Floating-Point Formalizaton in HOL
  Light}}.
\newblock \bibinfo{journal}{{\em Electronic Notes in Theoretical Computer
  Science\/}}  \bibinfo{volume}{317} (\bibinfo{year}{2015}),
  \bibinfo{pages}{101--107}.
\newblock


\bibitem[\protect\citeauthoryear{Jeannet and Min{\'e}}{Jeannet and
  Min{\'e}}{2009}]%
        {Jeannet2009}
\bibfield{author}{\bibinfo{person}{B.~ Jeannet} {and} \bibinfo{person}{A.~
  Min{\'e}}.} \bibinfo{year}{2009}\natexlab{}.
\newblock \showarticletitle{{Apron: A Library of Numerical Abstract Domains for
  Static Analysis}}. In \bibinfo{booktitle}{{\em CAV}}.
\newblock


\bibitem[\protect\citeauthoryear{Kinsman and Nicolici}{Kinsman and
  Nicolici}{2009}]%
        {Kinsman2009}
\bibfield{author}{\bibinfo{person}{A.~B.~ Kinsman} {and} \bibinfo{person}{N.~
  Nicolici}.} \bibinfo{year}{2009}\natexlab{}.
\newblock \showarticletitle{{Finite Precision Bit-Width Allocation using
  SAT-Modulo Theory}}. In \bibinfo{booktitle}{{\em DATE}}.
\newblock


\bibitem[\protect\citeauthoryear{Lam and Hollingsworth}{Lam and
  Hollingsworth}{2016}]%
        {Lam2016}
\bibfield{author}{\bibinfo{person}{M.~O.~ Lam} {and} \bibinfo{person}{J.~K.~
  Hollingsworth}.} \bibinfo{year}{2016}\natexlab{}.
\newblock \showarticletitle{{Fine-grained floating-point precision analysis}}.
\newblock \bibinfo{journal}{{\em Intl. Journal of High Performance Computing
  Applications\/}} (\bibinfo{date}{June} \bibinfo{year}{2016}).
\newblock


\bibitem[\protect\citeauthoryear{Lam, Hollingsworth, de~Supinski, and
  Legendre}{Lam et~al\mbox{.}}{2013}]%
        {Lam2013b}
\bibfield{author}{\bibinfo{person}{M.~O.~ Lam}, \bibinfo{person}{J.~K.~
  Hollingsworth}, \bibinfo{person}{B.~R.~ de Supinski}, {and}
  \bibinfo{person}{M.~P.~ Legendre}.} \bibinfo{year}{2013}\natexlab{}.
\newblock \showarticletitle{{Automatically Adapting Programs for
  Mixed-precision Floating-point Computation}}. In \bibinfo{booktitle}{{\em
  ICS}}.
\newblock


\bibitem[\protect\citeauthoryear{Lee, Gaffar, Cheung, Mencer, Luk, and
  Constantinides}{Lee et~al\mbox{.}}{2006}]%
        {Lee2006}
\bibfield{author}{\bibinfo{person}{D.~U.~ Lee}, \bibinfo{person}{A.~A.~
  Gaffar}, \bibinfo{person}{R.~C.~C.~ Cheung}, \bibinfo{person}{O.~ Mencer},
  \bibinfo{person}{W.~ Luk}, {and} \bibinfo{person}{G.~A.~ Constantinides}.}
  \bibinfo{year}{2006}\natexlab{}.
\newblock \showarticletitle{{Accuracy-Guaranteed Bit-Width Optimization}}.
\newblock \bibinfo{journal}{{\em Trans. Comp.-Aided Des. Integ. Cir. Sys.\/}}
  \bibinfo{volume}{25}, \bibinfo{number}{10} (\bibinfo{year}{2006}),
  \bibinfo{pages}{1990--2000}.
\newblock


\bibitem[\protect\citeauthoryear{Magron, Constantinides, and Donaldson}{Magron
  et~al\mbox{.}}{2015}]%
        {Magron2015}
\bibfield{author}{\bibinfo{person}{V.~ Magron}, \bibinfo{person}{G.~A.~
  Constantinides}, {and} \bibinfo{person}{A.~F.~ Donaldson}.}
  \bibinfo{year}{2015}\natexlab{}.
\newblock \showarticletitle{{Certified Roundoff Error Bounds Using Semidefinite
  Programming}}.
\newblock \bibinfo{journal}{{\em CoRR\/}}  \bibinfo{volume}{abs/1507.03331}
  (\bibinfo{year}{2015}).
\newblock


\bibitem[\protect\citeauthoryear{Majumdar, Saha, and Zamani}{Majumdar
  et~al\mbox{.}}{2012}]%
        {Majumdar2012}
\bibfield{author}{\bibinfo{person}{R.~ Majumdar}, \bibinfo{person}{I.~ Saha},
  {and} \bibinfo{person}{M.~ Zamani}.} \bibinfo{year}{2012}\natexlab{}.
\newblock \showarticletitle{{Synthesis of Minimal-error Control Software}}. In
  \bibinfo{booktitle}{{\em EMSOFT}}.
\newblock


\bibitem[\protect\citeauthoryear{Mallik, Sinha, Banerjee, and Zhou}{Mallik
  et~al\mbox{.}}{2007}]%
        {Mallik2007}
\bibfield{author}{\bibinfo{person}{A.~ Mallik}, \bibinfo{person}{D.~ Sinha},
  \bibinfo{person}{P.~ Banerjee}, {and} \bibinfo{person}{H.~ Zhou}.}
  \bibinfo{year}{2007}\natexlab{}.
\newblock \showarticletitle{{Low-Power Optimization by Smart Bit-Width
  Allocation in a SystemC-Based ASIC Design Environment.}}
\newblock \bibinfo{journal}{{\em IEEE Trans. on CAD of Integrated Circuits and
  Systems\/}} (\bibinfo{year}{2007}).
\newblock


\bibitem[\protect\citeauthoryear{Moore}{Moore}{1966}]%
        {Moore1966}
\bibfield{author}{\bibinfo{person}{R.~ Moore}.}
  \bibinfo{year}{1966}\natexlab{}.
\newblock \bibinfo{booktitle}{{\em Interval {A}nalysis}}.
\newblock \bibinfo{publisher}{Prentice-Hall}.
\newblock


\bibitem[\protect\citeauthoryear{Osborne, Cheung, Coutinho, Luk, and
  Mencer}{Osborne et~al\mbox{.}}{2007}]%
        {Osborne2007}
\bibfield{author}{\bibinfo{person}{W.~G.~ Osborne}, \bibinfo{person}{R.~C.~C.~
  Cheung}, \bibinfo{person}{J.~ Coutinho}, \bibinfo{person}{W.~ Luk}, {and}
  \bibinfo{person}{O.~ Mencer}.} \bibinfo{year}{2007}\natexlab{}.
\newblock \showarticletitle{{Automatic Accuracy-Guaranteed Bit-Width
  Optimization for Fixed and Floating-Point Systems}}. In
  \bibinfo{booktitle}{{\em Field Programmable Logic and Applications}}.
  \bibinfo{pages}{617--620}.
\newblock


\bibitem[\protect\citeauthoryear{Panchekha, Sanchez-Stern, Wilcox, and
  Tatlock}{Panchekha et~al\mbox{.}}{2015}]%
        {Panchekha2015}
\bibfield{author}{\bibinfo{person}{P.~ Panchekha}, \bibinfo{person}{A.~
  Sanchez-Stern}, \bibinfo{person}{J.~R.~ Wilcox}, {and} \bibinfo{person}{Z.~
  Tatlock}.} \bibinfo{year}{2015}\natexlab{}.
\newblock \showarticletitle{{Automatically Improving Accuracy for Floating
  Point Expressions}}. In \bibinfo{booktitle}{{\em PLDI}}.
\newblock


\bibitem[\protect\citeauthoryear{Pang, Radecka, and Zilic}{Pang
  et~al\mbox{.}}{2011}]%
        {Pang2011}
\bibfield{author}{\bibinfo{person}{Y.~ Pang}, \bibinfo{person}{K.~ Radecka},
  {and} \bibinfo{person}{Z.~ Zilic}.} \bibinfo{year}{2011}\natexlab{}.
\newblock \showarticletitle{{An Efficient Hybrid Engine to Perform Range
  Analysis and Allocate Integer Bit-widths for Arithmetic Circuits}}. In
  \bibinfo{booktitle}{{\em ASPDAC}}.
\newblock


\bibitem[\protect\citeauthoryear{Poli, Langdon, and McPhee}{Poli
  et~al\mbox{.}}{2008}]%
        {Poli08}
\bibfield{author}{\bibinfo{person}{R.~ Poli}, \bibinfo{person}{W.~B.~ Langdon},
  {and} \bibinfo{person}{N.~F.~ McPhee}.} \bibinfo{year}{2008}\natexlab{}.
\newblock \bibinfo{booktitle}{{\em {A} {F}ield {G}uide to {G}enetic
  {P}rogramming}}.
\newblock \bibinfo{publisher}{Lulu Enterprises}.
\newblock


\bibitem[\protect\citeauthoryear{Prokopec}{Prokopec}{2012}]%
        {Prokopec2012}
\bibfield{author}{\bibinfo{person}{A.~ Prokopec}.}
  \bibinfo{year}{2012}\natexlab{}.
\newblock \bibinfo{title}{ScalaMeter}.
\newblock \bibinfo{howpublished}{\url{https://scalameter.github.io/}}.
  (\bibinfo{year}{2012}).
\newblock


\bibitem[\protect\citeauthoryear{P\"{u}schel, Moura, Singer, Xiong, Johnson,
  Padua, Veloso, and Johnson}{P\"{u}schel et~al\mbox{.}}{2004}]%
        {Puschel2004}
\bibfield{author}{\bibinfo{person}{M.~ P\"{u}schel}, \bibinfo{person}{J.~M.~F.~
  Moura}, \bibinfo{person}{B.~ Singer}, \bibinfo{person}{J.~ Xiong},
  \bibinfo{person}{J.~R.~ Johnson}, \bibinfo{person}{D.~A.~ Padua},
  \bibinfo{person}{M.~M.~ Veloso}, {and} \bibinfo{person}{R.~W.~ Johnson}.}
  \bibinfo{year}{2004}\natexlab{}.
\newblock \showarticletitle{{Spiral - A Generator for Platform-Adapted
  Libraries of Signal Processing Alogorithms.}}
\newblock \bibinfo{journal}{{\em IJHPCA\/}} \bibinfo{volume}{18},
  \bibinfo{number}{1} (\bibinfo{year}{2004}), \bibinfo{pages}{21--45}.
\newblock


\bibitem[\protect\citeauthoryear{Ramananandro, Mountcastle, Meister, and
  Lethin}{Ramananandro et~al\mbox{.}}{2016}]%
        {Ramananandro2016}
\bibfield{author}{\bibinfo{person}{T.~ Ramananandro}, \bibinfo{person}{P.~
  Mountcastle}, \bibinfo{person}{B.~ Meister}, {and} \bibinfo{person}{R.~
  Lethin}.} \bibinfo{year}{2016}\natexlab{}.
\newblock \showarticletitle{{A Unified Coq Framework for Verifying C Programs
  with Floating-Point Computations}}. In \bibinfo{booktitle}{{\em CPP}}.
\newblock


\bibitem[\protect\citeauthoryear{Rubio-Gonz\'{a}les, Nguyen, Mehne, Sen,
  Demmel, Kahan, Iancu, Lavrijsen, Bailey, and Hough}{Rubio-Gonz\'{a}les
  et~al\mbox{.}}{2016}]%
        {Rubio-Gonzales2016}
\bibfield{author}{\bibinfo{person}{C.~ Rubio-Gonz\'{a}les},
  \bibinfo{person}{C.~ Nguyen}, \bibinfo{person}{B.~ Mehne},
  \bibinfo{person}{K.~ Sen}, \bibinfo{person}{J.~ Demmel}, \bibinfo{person}{W.~
  Kahan}, \bibinfo{person}{C.~ Iancu}, \bibinfo{person}{W.~ Lavrijsen},
  \bibinfo{person}{D.~H.~ Bailey}, {and} \bibinfo{person}{D.~ Hough}.}
  \bibinfo{year}{2016}\natexlab{}.
\newblock \showarticletitle{{Floating-Point Precision Tuning Using Blame
  Analysis}}. In \bibinfo{booktitle}{{\em ICSE}}.
\newblock


\bibitem[\protect\citeauthoryear{Rubio-Gonz{\'a}lez, Nguyen, Nguyen, Demmel,
  Kahan, Sen, Bailey, Iancu, and Hough}{Rubio-Gonz{\'a}lez
  et~al\mbox{.}}{2013}]%
        {Rubio-Gonzalez2013}
\bibfield{author}{\bibinfo{person}{C.~ Rubio-Gonz{\'a}lez},
  \bibinfo{person}{C.~ Nguyen}, \bibinfo{person}{H.~D.~ Nguyen},
  \bibinfo{person}{J.~ Demmel}, \bibinfo{person}{W.~ Kahan},
  \bibinfo{person}{K.~ Sen}, \bibinfo{person}{D.~H.~ Bailey},
  \bibinfo{person}{C.~ Iancu}, {and} \bibinfo{person}{D.~ Hough}.}
  \bibinfo{year}{2013}\natexlab{}.
\newblock \showarticletitle{{Precimonious: Tuning Assistant for Floating-point
  Precision}}. In \bibinfo{booktitle}{{\em SC}}.
\newblock


\bibitem[\protect\citeauthoryear{R{\"u}mmer and Wahl}{R{\"u}mmer and
  Wahl}{2010}]%
        {Rummer2010}
\bibfield{author}{\bibinfo{person}{P.~ R{\"u}mmer} {and} \bibinfo{person}{T.~
  Wahl}.} \bibinfo{year}{2010}\natexlab{}.
\newblock \showarticletitle{An {SMT-LIB} {T}heory of {B}inary
  {F}loating-{P}oint {A}rithmetic}. In \bibinfo{booktitle}{{\em SMT}}.
\newblock


\bibitem[\protect\citeauthoryear{Schkufza, Sharma, and Aiken}{Schkufza
  et~al\mbox{.}}{2014}]%
        {Schkufza2014}
\bibfield{author}{\bibinfo{person}{E.~ Schkufza}, \bibinfo{person}{R.~ Sharma},
  {and} \bibinfo{person}{A.~ Aiken}.} \bibinfo{year}{2014}\natexlab{}.
\newblock \showarticletitle{{Stochastic Optimization of Floating-point Programs
  with Tunable Precision}}. In \bibinfo{booktitle}{{\em PLDI}}.
\newblock


\bibitem[\protect\citeauthoryear{Solovyev, Jacobsen, Rakamaric, and
  Gopalakrishnan}{Solovyev et~al\mbox{.}}{2015}]%
        {Solovyev2015}
\bibfield{author}{\bibinfo{person}{A.~ Solovyev}, \bibinfo{person}{C.~
  Jacobsen}, \bibinfo{person}{Z.~ Rakamaric}, {and} \bibinfo{person}{G.~
  Gopalakrishnan}.} \bibinfo{year}{2015}\natexlab{}.
\newblock \showarticletitle{{Rigorous Estimation of Floating-Point Round-off
  Errors with Symbolic Taylor Expansions}}. In \bibinfo{booktitle}{{\em FM}}.
\newblock


\bibitem[\protect\citeauthoryear{Vuduc, Demmel, and Bilmes}{Vuduc
  et~al\mbox{.}}{2004}]%
        {Vuduc2004}
\bibfield{author}{\bibinfo{person}{R.~ Vuduc}, \bibinfo{person}{J.~W.~ Demmel},
  {and} \bibinfo{person}{J.~A.~ Bilmes}.} \bibinfo{year}{2004}\natexlab{}.
\newblock \showarticletitle{Statistical Models for Empirical Search-Based
  Performance Tuning}.
\newblock \bibinfo{journal}{{\em Int. J. High Perform. Comput. Appl.\/}}
  \bibinfo{volume}{18}, \bibinfo{number}{1} (\bibinfo{date}{Feb.}
  \bibinfo{year}{2004}), \bibinfo{pages}{65--94}.
\newblock


\bibitem[\protect\citeauthoryear{Zeller and Hildebrandt}{Zeller and
  Hildebrandt}{2002}]%
        {Zeller2002}
\bibfield{author}{\bibinfo{person}{A.~ Zeller} {and} \bibinfo{person}{R.~
  Hildebrandt}.} \bibinfo{year}{2002}\natexlab{}.
\newblock \showarticletitle{{Simplifying and Isolating Failure-Inducing
  Input.}}
\newblock \bibinfo{journal}{{\em IEEE Trans. Software Eng.\/}}
  \bibinfo{volume}{28}, \bibinfo{number}{2} (\bibinfo{year}{2002}),
  \bibinfo{pages}{183--200}.
\newblock


\end{thebibliography}

\end{document}